\documentclass[review]{elsarticle}

\usepackage{lineno,hyperref,amsmath,amssymb,bm}
\usepackage{subfigure}

\journal{Optics and Lasers in Engineering}

\begin{document}
	
	\begin{frontmatter}
		
		\title{Performance of fiber-based QAM/FSO systems in turbulence with anisotropic tilt angle and random angular jitter}
		
		\author[label1,label2]{Chao Zhai\corref{cor1}}
		\cortext[cor1]{Corresponding author.}
		\ead{zhaichao@qhd.neu.edu.cn}
		
		\author[label1,label2]{Zhenyuan Xue}
		
		\affiliation[label1]{organization={School of Computer and Communication Engineering}, addressline={Northeastern University at Qinhuangdao}, city={Qinhuangdao}, postcode={066004}, country={China}}
		
		\affiliation[label2]{organization={Hebei Key Laboratory of Marine Perception Network and Data Processing}, addressline={Northeastern University at Qinhuangdao}, city={Qinhuangdao}, postcode={066004}, country={China}}
		
		\begin{abstract}
			Nowadays, the subsistent anisotropic non-Kolmogorov (ANK) turbulence models are all established on the supposition that the long axis of turbulence cell ought to be level with the ground. Nevertheless, Beason et al. and Wang et al. have illustrated through their recent experimental results that there is an anisotropic tilt angle in the turbulence cell, i.e., the long axis of turbulence cell is probably not level with the ground but has a particular angle with the ground. For the practical free-space optical (FSO) communication link, bias error and random angular jitter are critical elements which influence the fiber-based FSO communication system performance. In this paper, employing the new ANK turbulence spectrum models in the horizontal link with anisotropic tilt angle, we derive the normalized probability density function (PDF) for Gamma-Gamma distribution with anisotropic tilt angle, and the PDF of fiber-coupling efficiency in the presence of bias error, anisotropic tilt angle, and random angular jitter. And then the average bit error rate (BER) expression of fiber-based $J \times Q$ rectangular quadrature amplitude modulation (QAM) FSO systems for a plane wave transmission through the weak ANK horizontal link in the presence of bias error, anisotropic tilt angle, and random angular jitter is developed.
		\end{abstract}
		
		
	\end{frontmatter}

\section{Introduction}
FSO communication technology has the advantages of large capacity, high confidentiality, and strong anti-interference ability, which is one of the pivotal technologies to realize the high-speed real-time space-ground integrated network \cite{Khamees2023,Wang2023,Wang2024,Yu2025}. Utilizing mature ground optical fiber communication technology to improve the FSO communication system performance has turned into one of the vital options, wavelength division multiplexing technology and optical amplifier can successfully enhance the communication data rate and the detection sensitivity for the FSO communication systems, and it is very easy to connect FSO communications with terrestrial fiber networks through fiber-based FSO communication systems. The study on fiber-based FSO communication systems has turned into a research hotspot. Zhang et al. proposed a novel fiber-based FSO coherent receiver with a transmission rate of 22.4 Gbit/s for inter-satellite communication, which took advantage of established fiber-optic components and utilized the fine-pointing subsystem installed in FSO terminals to minimize the influence of satellite platform vibrations \cite{Zhang2013}. On the basis of few-mode-fiber coupling, Wang et al. proposed and verified two simplified mode diversity coherent detection receivers for FSO communication systems under moderate-to-strong turbulence \cite{Wang2022}. Trinh et al. experimentally investigated the PDF of coupling efficiency for a 200 $ \mu $m multi-mode fiber using data from an FSO downlink from the small optical link for international space station terminal to a 40 cm sub-aperture optical ground station supported by a fine-tracking system \cite{Trinh2023}. We developed the temporal power spectrum of irradiance fluctuations and the PDF of fiber-coupling efficiency for a plane wave transmitting through the weak ANK satellite-to-ground downlink \cite{Zhai2021a}. Furthermore, QAM is an exhilarating technique to achieve a high communication data rate without improving the system bandwidth, which can definitely significantly enhance the FSO communication system performance. So far there are still few studies on the fiber-based QAM/FSO systems over atmospheric turbulence channels. It is imperative to develop the investigation for fiber-based FSO communication systems to the quadrature amplitude modulation.

Atmospheric turbulence is one of the most vital elements that restrict the FSO communication system performance. Even though the ANK turbulence model is presently the most rifely applied atmospheric turbulence model, Beason et al. and Wang et al. have illustrated clearly through their recent experimental results that the turbulence cell has an anisotropic tilt angle, i.e., the long axis of turbulence cell may not be level with the ground but has a particular angle with the ground \cite{Wang2017,Beason2018}. Subsequently, on the basis of ellipsoid hypothesis for turbulence cell, we have developed the ANK turbulence spectra with anisotropic tilt angle in the satellite and horizontal links, and the influences of azimuth angle and anisotropic tilt angle on the scintillation index have been analyzed \cite{Zhai2021b,Zhai2022a}. The study on optical wave transmission through the ANK turbulence with anisotropic tilt angle has turned into a research hotspot. Lin et al. developed the expression of scintillation index for a spherical wave transmitting over anisotropic oceanic turbulence with anisotropic tilt angle in the vertical link \cite{Lin2022}. Yu et al. developed expressions for the spatial coherence radius and the detection probability of the orbital angular momentum states of the Kummer beam in turbulence with an anisotropic tilt angle \cite{Yu2023}. We can find that the ANK turbulence model with anisotropic tilt angle is more closer to the practical situation for atmospheric turbulence. And in the actual process of optical wave atmospheric propagation, the variation of anisotropic tilt angle will influence the state of atmospheric turbulence, make the received light field change, and finally bring down the quality of the received laser beam. Nevertheless, there are still few studies on FSO communication systems and optical wave transmission in the ANK turbulence with anisotropic tilt angle. It is imperative to develop the investigation of fiber-based QAM/FSO systems in turbulence to the ANK turbulence with anisotropic tilt angle. 

For the practical FSO communication link, bias error and random angular jitter are generated by the unpredictability between the instantaneous direction of received laser and the nominal axis of fiber, which have a great effect on the fiber-coupling efficiency, and then greatly influence the fiber-based FSO communication system performance \cite{Toyoshima2006,Ma2009,Chen1989,Zhai2020}. Hence, we should know that the fiber-coupling efficiency is the combined result of atmospheric turbulence and random angular jitter. However, the impact of random angular jitter on the fiber-based FSO communication systems has not been considered for the most studies.

In this paper, we focus on the BER performance of fiber-based $J \times Q$ rectangular QAM/FSO communication systems for the weak ANK horizontal link in the presence of bias error, anisotropic tilt angle, and random angular jitter. The influences of anisotropic tilt angle in the atmospheric channel and random angular jitter in the FSO receiver system are both considered. To help the optimization of system design in engineering, the impacts of atmospheric turbulence parameters and communication system parameters on the BER performance in the horizontal link are numerically evaluated.

\section{System and atmospheric channel model}
During this part, a succinct clarification about modulation and demodulation procedures is exhibited, and then the atmospheric channel model is provided. Firstly, we consider a fiber-based FSO communication system employing arbitrary $J \times Q$ rectangular QAM in this paper. For the transmitter system, each block of ${\log _2}\left( {J \times Q} \right)$ data bits is modulated by an electrical QAM modulator, where $J$ represents the dimension of in-phase signals, and $Q$ represents the dimension of quadrature signals. The signal at the electrical QAM modulator output can be expressed as
\begin{equation}
	s\left( t \right) = {A_J}g\left( t \right)\cos 2\pi {f_c}t - {A_Q}g\left( t \right)\sin 2\pi {f_c}t,\hspace*{1em}0 \le t < {T_s},
	\label{eq:refname1}
\end{equation}
where $g\left( t \right)$ represents the signal pulse, $T_s$ represents the symbol time duration, $f_c$ represents the carrier frequency, ${R_s} = 1/{T_s}$ represents the symbol rate, $A_J$ represents the amplitude of in-phase signals, and $A_Q$ represents the amplitude of quadrature signals. We can employ the electrical QAM signal $s\left( t \right)$ to modulate the intensity of transmitter laser source. To prevent negative input, the electrical QAM signal $s\left( t \right)$ is DC-level shifted, and the transmitted optical signal has the subsequent expression:
\begin{equation}
	P\left( t \right) = {P_t}\left[ {1 + ms\left( t \right)} \right],
	\label{eq:refname2}
\end{equation}
where $P_t$ represents the transmitted optical power, and $m$ represents the modulation index (it is commonly supposed that $m = 1$). Subsequently, the optical signal is propagated through the atmospheric turbulence channel. For the receiver system, after coupling the spatial signal light into a single-mode fiber, accomplishing direct detection, and removing DC bias, we can convert the optical signal into an electrical one by use of a PIN photodiode, which can be expressed as \cite{Zhai2022}
\begin{equation}
	r\left( t \right) = R\eta I{P_r}s\left( t \right) + n\left( t \right),
	\label{eq:refname3}
\end{equation}
where $I$ represents the atmospheric turbulence-induced fading which can be simulated by Gamma-Gamma distribution for the weak-to-strong fluctuation regime, $\eta$ represents the fiber-coupling efficiency, $R$ represents the photodiode responsivity, ${P_r} = {a_l}{D^2}{P_t}/{\vartheta ^2}{L^2}$ represents the received optical power without atmospheric turbulence, $L$ represents the link distance, $D$ represents the receiver diameter, $\vartheta$ represents the divergence angle, and $a_l$ represents the link energy loss. Due to the utilization of PIN photodiode, the thermal noise $n\left( t \right)$ is dominant for the total receiver noise, and we can 
simulate it by utilizing a zero-mean additive white Gaussian noise (AWGN) with variance $\sigma _n^2 = {N_0}/2$ and $N_0$ being the noise power spectrum density \cite{Djordjevic2016}.

By use of Eq.~(\ref{eq:refname3}), the instantaneous electrical signal-to-noise ratio (SNR) has the subsequent expression \cite{Zhai2022}:
\begin{equation}
	{\gamma _s}  = \frac{{{R^2}{\eta ^2}{I^2}P_r^2}}{{2\sigma _n^2}}.
	\label{eq:refname4}
\end{equation}
To support the BER estimation of $J \times Q$ rectangular QAM, we commonly utilize the SNR per bit, which can be expressed as \cite{Proakis2008}
\begin{equation}
	{\gamma _b} = \frac{1}{{{R_s}{{\log }_2}\left( {J \cdot Q} \right)}}\frac{{{R^2}{\eta ^2}{I^2}P_r^2}}{{2\sigma _n^2}}.
	\label{eq:refname5}
\end{equation}

\begin{figure}[ht!]
	\centering
	\includegraphics[width=9cm]{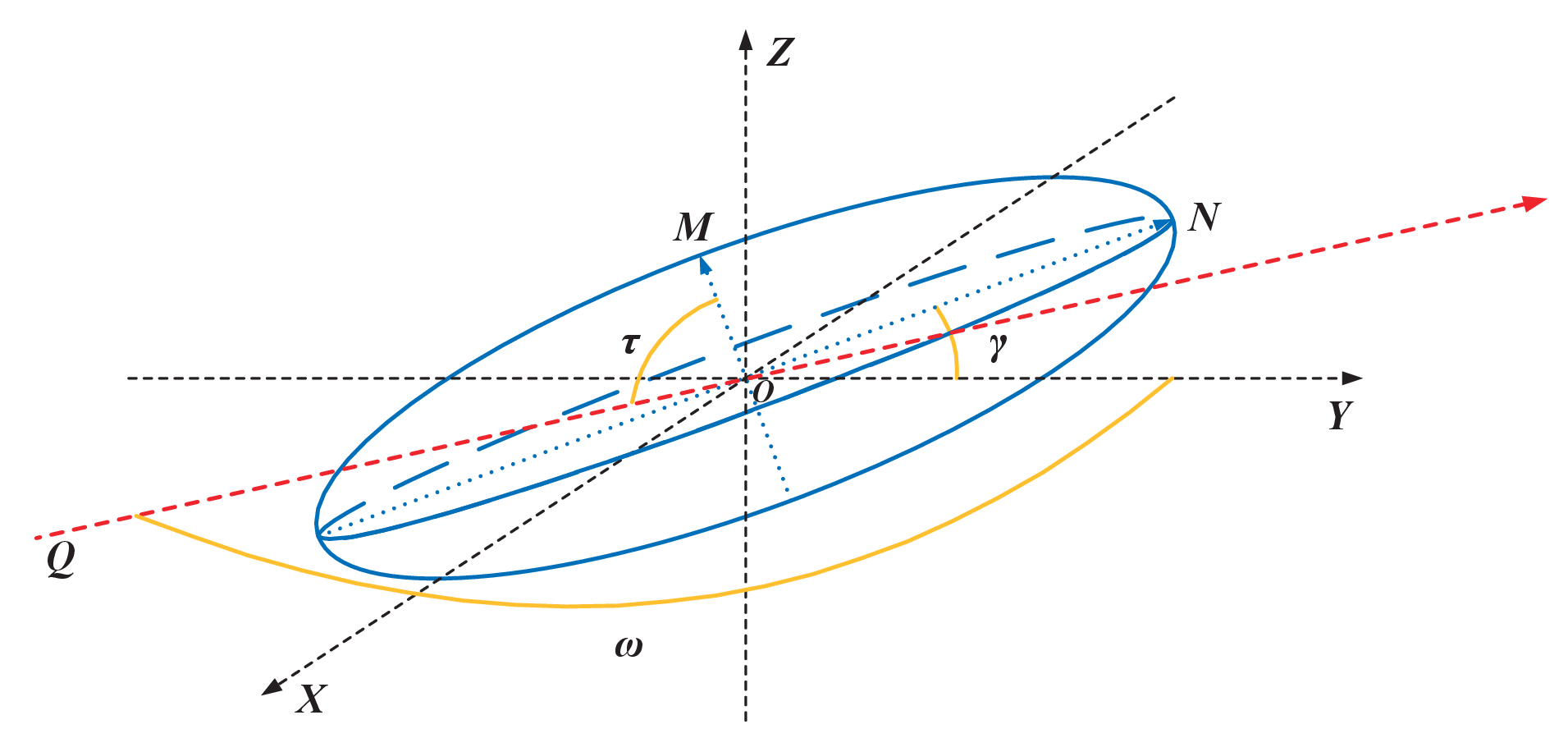}
	\caption{The anisotropic turbulence cell for the horizontal link with anisotropic tilt angle.}
	\label{Fig01}
\end{figure}

Nextly, the atmospheric channel model with anisotropic tilt angle $\gamma$ for the ANK horizontal link will be developed. As stated in the introduction and our previous research \cite{Zhai2021b}, the turbulence cell has an anisotropic tilt angle, i.e., the long axis of turbulence cell may not be level with the ground but has a particular angle with the ground \cite{Wang2017,Beason2018}. In Fig.~\ref{Fig01}, the anisotropic tilt angle $\gamma$ represents the angle between the plane where the long axes of ellipsoid are located and the horizontal plane, counterclockwise is positive, and the value range is 0 $ ^\circ $ to 180 $ ^\circ $. The azimuth angle $\omega$ for the horizontal link represents the angle between the propagation direction of optical wave $QO$ and the positive half of the $y$-axis, clockwise is positive, the value range is 0 $ ^\circ $ to 360 $ ^\circ $. And $\tau  = \arccos \left( { - \sin \gamma \cos \omega } \right)$ represents the angle between the short axis of ellipsoid and the propagation direction of optical wave $QO$, as displayed in Fig.~\ref{Fig01}. We can simply find from Fig.~\ref{Fig01} that owing to the anisotropic tilt angle, the turbulence cell rotates counterclockwise around the $x$-axis by an angle $\gamma$. As exhibited in Fig.~\ref{Fig01}, $QO$ represents the propagation direction of optical wave in the horizontal link, the horizontal plane is the plane $XOY$, and when the anisotropic tilt angle $\gamma$ is not 0 $ ^\circ $ and 180 $ ^\circ $, the radius of curvature for the interface between the anisotropic turbulence cell and the optical propagation beam varies with the changing of the azimuth angle $\omega$. Furthermore, because the long axis length of the turbulence cell is usually one to several times the short axis length, we have $ON = \mu s$, $OM = s$, where $\mu$ represents the anisotropic factor that is the ratio of long axis to short axis for the turbulence cell. Therefore, the anisotropic turbulence cell can be assumed to be an ellipsoid, consistent with many studies on anisotropic turbulence \cite{Zhai2022a,Lin2022,Cui2015a}. Note that in the actual process of laser communication, the variations of anisotropic tilt angle $\gamma$ and the azimuth angle $\omega$ will influence the state of atmospheric turbulence, make the received signal light field change, and finally bring down the performance of space laser communication.

On the basis of the turbulence cell anisotropy model with anisotropic tilt angle $\gamma$ in the horizontal link that we have established \cite{Zhai2021b}, the anisotropic factors $\mu _x$ and $\mu _y$ for the horizontal link can be expressed as
\begin{equation}
	{\mu _x} = \sqrt {\frac{{{\mu ^2} + {{\tan }^2}\tau }}{{1 + {{\tan }^2}\tau }}},\hspace*{1em}{\mu _y} = \sqrt {\frac{{{\mu ^2} + {{\tan }^2}\tau }}{{1 + {\mu ^2}{{\tan }^2}\tau }}},
	\label{eq:refname6}
\end{equation}

Substituting Eq.~(\ref{eq:refname6}) into the subsistent ANK turbulence spectrum models \cite{Cui2015a}, the power spectrum models with anisotropic tilt angle $\gamma$ for the ANK horizontal link have the subsequent expressions:
\begin{equation}
	{\Phi _n}\left( {\kappa ,\alpha ,{\mu _x},{\mu _y}} \right) = \frac{{A\left( \alpha  \right)\tilde C_n^2{\mu _x}{\mu _y}}}{{{{\left( {\mu _x^2\kappa _x^2 + \mu _y^2\kappa _y^2 + \kappa _z^2} \right)}^{\alpha /2}}}},\hspace*{1em}1/{L_0} < \kappa  < 1/{l_0},
	\label{eq:refname7}
\end{equation}
\begin{equation}
	{\Phi _n}\left( {\kappa ,\alpha ,{\mu _x},{\mu _y}} \right) = \frac{{A\left( \alpha  \right)\tilde C_n^2{\mu _x}{\mu _y}}}{{{{\left( {\mu _x^2\kappa _x^2 + \mu _y^2\kappa _y^2 + \kappa _z^2 + \kappa _0^{\rm{2}}} \right)}^{\alpha /2}}}}\exp \left( { - \frac{{\mu _x^2\kappa _x^2 + \mu _y^2\kappa _y^2 + \kappa _z^2}}{{\kappa _l^2}}} \right),\hspace*{0.2em}3 < \alpha  < 4,
	\label{eq:refname8}
\end{equation}
where $\kappa  = \sqrt {\mu _x^2\kappa _x^2 + \mu _y^2\kappa _y^2 + \kappa _z^2} $, $\kappa _x$, $\kappa _y$, and $\kappa _z$ represent the $x$, $y$, and $z$ components of $\kappa$, $\kappa$ represents the spatial wavenumber vector, $\mu _x$ and $\mu _y$ represent the anisotropic factors which are able to demonstrate the anisotropy of turbulence cells, $\alpha$ represents the spectral power law, $\tilde C_n^2$ represents the structure constant of ANK turbulence in units of m$ ^{3 - \alpha } $, $ \kappa_l = c\left( \alpha  \right)/l_0 $, ${\kappa _0} = 2\pi /{L_0}$, $ l_0 $ and $ L_0 $ represent the inner and outer scales of turbulence, $ A\left( \alpha  \right) = \Gamma \left( {\alpha  - 1} \right)\cos \left( {\alpha \pi /2} \right)/4{\pi ^2} $, $ \Gamma \left( x \right) $ represents the Gamma function, and $ c\left( \alpha  \right) = {\left[ {\left( {2\pi /3} \right)\Gamma \left( {2.5 - 0.5\alpha } \right)A\left( \alpha  \right)} \right]^{1/\left( {\alpha  - 5} \right)}} $.

\begin{figure}[ht!]
	\centering
	\includegraphics[width=9cm]{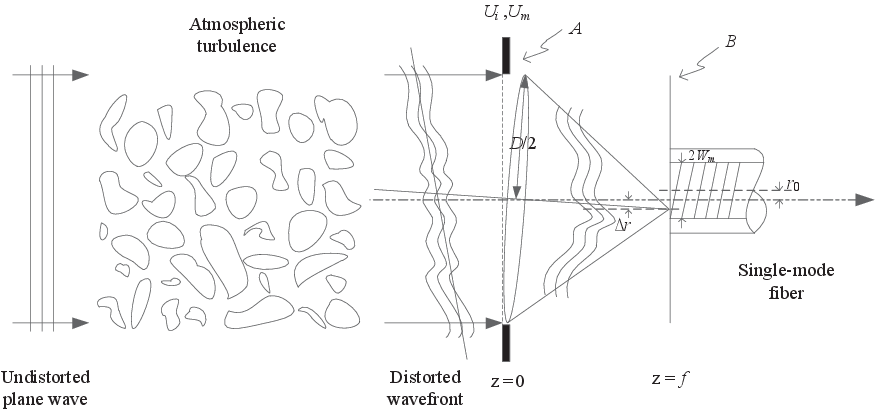}
	\caption{Coupling geometry in the presence of bias error and random angular jitter.}
	\label{Fig02}
\end{figure}

For the practical FSO communication link, bias error and random angular jitter are generated by the unpredictability between the instantaneous direction of received laser and the nominal axis of fiber, which have a great effect on the fiber-coupling efficiency, and then greatly influence the fiber-based FSO communication system performance \cite{Toyoshima2006,Ma2009,Chen1989,Zhai2020}. Hence, we can deduce that the fiber-coupling efficiency is the combined result of atmospheric turbulence and random angular jitter. As shown in Fig.~\ref{Fig02}, the coupling geometry in the presence of bias error and random angular jitter through an atmospheric channel is given. By use of a thin diffraction-limited lens located at the receiver aperture plane $A$, the incident optical field distorted by atmospheric turbulence is focused on the single-mode fiber end which is located at the focal plane $B$, and $ f $ represents the focal length.

The definition of fiber-coupling efficiency for an optical wave is the ratio between the power coupled into the fiber $P_c$ and the power available on the receiver plane $P_a$, which can be expressed as \cite{Zhai2021a}
\begin{equation}
	\eta  = \frac{{{P_c}}}{{{P_a}}} = \frac{{{{\left| {\int_A {{U_i}\left( \bf{r} \right)U_m^ * \left( \bf{r} \right)d\bf{r}} } \right|}^2}}}{{\int_A {{{\left| {{U_i}\left( \bf{r} \right)} \right|}^2}d\bf{r}} }},
	\label{eq:refname9}
\end{equation}
where $ {{U_m}\left( \bf{r} \right)} $ and $ {{U_i}\left( \bf{r} \right)} $ represent severally the normalized backpropagated fiber-mode profile and the incident optical field at the receiver plane $A$, and the * represents the complex conjugate. The incident optical field of a plane wave at the receiver plane has the subsequent expression:
\begin{equation}
	{U_i}\left( {\bf{r}} \right) = {A_i}\exp \left[ { - i\phi \left( {\bf{r}} \right)} \right],
	\label{eq:refname10}
\end{equation}
where ${\phi \left( {\bf{r}} \right)}$ represents the phase perturbation owing to atmospheric turbulence, and $A_i$ represents the optical amplitude. Since the phase perturbation is the completely dominant factor affecting the fiber-coupling efficiency, the amplitude perturbation can be ignored \cite{Zhai2021a}.

In the presence of random angular jitter, since the incident optical field $ {{U_i}\left( \bf{r} \right)} $ is tilted by the random jitter angle $\varepsilon$, the focused optical field will generate a lateral displacement $\Delta r = \varepsilon f$ at the focal plane. Then we can equivalent the effect caused by random angular jitter to a lateral displacement $ \Delta r $ between the nominal axis of fiber and the optical axis of receiver lens at the focal plane. When a lateral displacement $ \Delta r $ exists at the focal plane, the fiber-mode profile backpropagated to the front surface of receiver lens has the subsequent expression \cite{Ma2009}:
\begin{equation}
	{U_m}\left( {\bf{r}} \right) = \frac{{\sqrt {2\pi } {W_m}}}{{\lambda f}}\exp \left[ { - {{\left( {\frac{{\pi {W_m}}}{{\lambda f}}} \right)}^2}{r^2}} \right]\exp \left[ {i\frac{{2\pi }}{{\lambda f}}\cos \left( {\varphi  - \Omega } \right)r\Delta r} \right],
	\label{eq:refname11}
\end{equation}
where $ \lambda $ represents the wavelength, $ \Omega $ represents the angle between the $ x $ axis and $ \Delta {\bf{r}} $, $ W_m $ represents the fiber-mode field radius, and $ \varphi $ represents the angle between the $ x $ axis and $ \bf{r} $.

In comparison to the lateral displacement $ \Delta r $ generated by random angular jitter, the bias error $ r_0 $ represents the static radial displacement between the nominal axis of fiber and the optical axis of receiver lens at the focal plane. When both bias error and random angular jitter exist, we can obtain the distribution connected with $ \Delta r $ and $ r_0 $ by the Nakagami-Rice distribution \cite{Ma2009},
\begin{equation}
	f\left( {\Delta r,{r_0}} \right) = \frac{{\Delta r}}{{\sigma _e^2}}\exp \left( { - \frac{{\Delta {r^2} + r_0^2}}{{2\sigma _e^2}}} \right){I_0}\left( {\frac{{\Delta r \cdot {r_0}}}{{\sigma _e^2}}} \right),
	\label{eq:refname12}
\end{equation}
where $ {I_0}\left( x \right) $ represents the modified Bessel function for the zero order and first kind, $\sigma _\varepsilon$ represents the standard deviation for the random jitter angle $\varepsilon$, and ${\sigma _e} = {\sigma _\varepsilon }f$ represents the standard deviation for the lateral displacement $ \Delta r $.

Substituting Eqs.~(\ref{eq:refname10}) and~(\ref{eq:refname11}) into Eq.~(\ref{eq:refname9}), the fiber-coupling efficiency for a plane wave in the presence of bias error and random angular jitter can be expressed as
\begin{equation}
	\begin{array}{l}
		\eta  = \frac{{8W_m^2}}{{{\lambda ^2}{f^2}{D^2}}}{\left\{ {\int_A {\cos \left[ {\phi \left( {\bf{r}} \right) + \frac{{2\pi }}{{\lambda f}}\cos \left( {\varphi  - \Omega } \right)r\Delta r} \right]\exp \left[ { - {{\left( {\frac{{\pi {W_m}}}{{\lambda f}}} \right)}^2}{r^2}} \right]d{\bf{r}}} } \right\}^2}\\
		\hspace*{1.7em}+ \frac{{8W_m^2}}{{{\lambda ^2}{f^2}{D^2}}}{\left\{ {\int_A {\sin \left[ {\phi \left( {\bf{r}} \right) + \frac{{2\pi }}{{\lambda f}}\cos \left( {\varphi  - \Omega } \right)r\Delta r} \right]\exp \left[ { - {{\left( {\frac{{\pi {W_m}}}{{\lambda f}}} \right)}^2}{r^2}} \right]d{\bf{r}}} } \right\}^2}\\
		\hspace*{1em}= {\eta _0}\left( {a_r^2 + a_i^2} \right)\\
		\hspace*{1em}= {\eta _0}{a^2},
	\end{array}
	\label{eq:refname13}
\end{equation}
where
\begin{equation}
	{\eta _0} = \frac{{{\pi ^2}W_m^2{D^2}}}{{2{\lambda ^2}{f^2}}},
	\label{eq:refname14}
\end{equation}
\begin{equation}
	{a_r} = {\left( {\frac{\pi }{4}{D^2}} \right)^{ - 1}}\int_A {\exp \left[ { - {{\left( {\frac{{\pi {W_m}}}{{\lambda f}}} \right)}^2}{r^2}} \right]} \cos \left[ {\phi \left( {\bf{r}} \right) + \frac{{2\pi }}{{\lambda f}}\cos \left( {\varphi  - \Omega } \right)r\Delta r} \right]d{\bf{r}},
	\label{eq:refname15}
\end{equation}
\begin{equation}
	{a_i} = {\left( {\frac{\pi }{4}{D^2}} \right)^{ - 1}}\int_A {\exp \left[ { - {{\left( {\frac{{\pi {W_m}}}{{\lambda f}}} \right)}^2}{r^2}} \right]} \sin \left[ {\phi \left( {\bf{r}} \right) + \frac{{2\pi }}{{\lambda f}}\cos \left( {\varphi  - \Omega } \right)r\Delta r} \right]d{\bf{r}},
	\label{eq:refname16}
\end{equation}
are the normalized forms for the integrals. Since ${\phi \left( {\bf{r}} \right)}$ is a random variable, we cannot get the analytical expressions of the integrals in Eqs.~(\ref{eq:refname15}) and~(\ref{eq:refname16}), but we can simply utilize the PDF of $a^2$ to develop the PDF of fiber-coupling efficiency.

Based on the approximation in speckle statistics of imaging optics \cite{Cagigal1998}, we can assume that the receiver aperture is made up of numerous independent correlation units with diameter $c_0$. The number of units in the receiver aperture with diameter $D$ is nearly proportional to ${\left( {D/{c_0}} \right)^2}$. Thus, we can express the integrals in Eqs.~(\ref{eq:refname15}) and~(\ref{eq:refname16}) as the finite sums over $N$ statistically independent units in the receiver aperture, which have the subsequent expressions:
\begin{equation}
	{a_r} = \frac{1}{N}\sum\limits_{n = 1}^N {\exp \left[ { - {{\left( {\frac{{\pi {W_m}}}{{\lambda f}}} \right)}^2}r_n^2} \right]} \cos \left[ {{\phi _n} + \frac{{2\pi }}{{\lambda f}}\cos \left( {{\varphi _n} - \Omega } \right){r_n}\Delta r} \right],
	\label{eq:refname17}
\end{equation}
\begin{equation}
	{a_i} = \frac{1}{N}\sum\limits_{n = 1}^N {\exp \left[ { - {{\left( {\frac{{\pi {W_m}}}{{\lambda f}}} \right)}^2}r_n^2} \right]} \sin \left[ {{\phi _n} + \frac{{2\pi }}{{\lambda f}}\cos \left( {{\varphi _n} - \Omega } \right){r_n}\Delta r} \right],
	\label{eq:refname18}
\end{equation}
where $c_0$ represents the Fried coherence length, $N$ represents the number of statistically independent units, for $D < {c_0}$, $N \approx 1$, and for $D > {c_0}$, $N \approx {\left( {D/{c_0}} \right)^2}$. The wave structure function of a plane wave for the weak Kolmogorov horizontal link has the subsequent expression \cite{Andrews2005}:
\begin{equation}
	D\left( \rho  \right) = 8{\pi ^2}{k^2}L\int_0^\infty  {\kappa {\Phi _n}\left( \kappa  \right)\left[ {1 - {J_0}\left( {\kappa \rho } \right)} \right]d\kappa },
	\label{eq:refname19}
\end{equation}
where $ J_0 (x) $ represents the Bessel function for the zero order and first kind, and $k = 2\pi /\lambda$.

Due to the Markov approximation which supposes that the refraction index is delta-correlated at any pair of points situated along the propagation direction, we can ignore the component $\kappa _z$ in Eq.~(\ref{eq:refname8}). Substituting Eq.~(\ref{eq:refname8}) into Eq.~(\ref{eq:refname19}) results in
\begin{equation}
	D\left( \rho  \right) = 4\pi {k^2}L\int_0^\infty  {\int_0^\infty  {{\Phi _n}\left( {\kappa ,\alpha ,{\mu _x},{\mu _y}} \right)} } \left[ {1 - {J_0}\left( {\kappa \rho } \right)} \right]d{\kappa _x}d{\kappa _y}.
	\label{eq:refname20}
\end{equation}

We can transform the anisotropic coordinate system of ${\Phi _n}\left( {\kappa ,\alpha ,{\mu _x},{\mu _y}} \right)$ in Eq.~(\ref{eq:refname8}) into an isotropic coordinate system with the help of subsequent mathematical substitutions,
\begin{equation}
	{\kappa _x} = \frac{{{q_x}}}{{{\mu _x}}} = \frac{{q\cos \theta }}{{{\mu _x}}},\hspace*{1em}{\kappa _y} = \frac{{{q_y}}}{{{\mu _y}}} = \frac{{q\sin \theta }}{{{\mu _y}}},\hspace*{1em}q = \sqrt {q_x^2 + q_y^2},
	\label{eq:refname21}
\end{equation}
\begin{equation}
	\kappa  = q\sqrt {\frac{{{{\cos }^2}\theta }}{{\mu _x^2}} + \frac{{{{\sin }^2}\theta }}{{\mu _y^2}}},\hspace*{1em}d{\kappa _x}d{\kappa _y} = \frac{{d{q_x}d{q_y}}}{{{\mu _x}{\mu _y}}} = \frac{{qdqd\theta }}{{{\mu _x}{\mu _y}}},
	\label{eq:refname22}
\end{equation}
\begin{equation}
	{\Phi _n}\left( {\kappa ,\alpha ,{\mu _x},{\mu _y}} \right) = A\left( \alpha  \right)\tilde C_n^2{\mu _x}{\mu _y}{\left( {{q^2} + \kappa _0^{\rm{2}}} \right)^{ - \frac{\alpha }{2}}}\exp \left( { - \frac{{{q^2}}}{{\kappa _l^2}}} \right).
	\label{eq:refname23}
\end{equation}

Inserting Eqs.~(\ref{eq:refname21}),~(\ref{eq:refname22}), and~(\ref{eq:refname23}) into Eq.~(\ref{eq:refname20}), we can get
\begin{equation}
	\begin{array}{l}
		D\left( \rho  \right) = 4\pi {k^2}LA\left( \alpha  \right)\tilde C_n^2\int_0^{2\pi } {\int_0^\infty  {q{{\left( {{q^2} + \kappa _0^{\rm{2}}} \right)}^{ - \frac{\alpha }{2}}}} } \\
		\hspace*{8em}\times \exp \left( { - \frac{{{q^2}}}{{\kappa _l^2}}} \right)\left[ {1 - {J_0}\left( {\rho q\sqrt {\frac{{{{\cos }^2}\theta }}{{\mu _x^2}} + \frac{{{{\sin }^2}\theta }}{{\mu _y^2}}} } \right)} \right]dqd\theta.
	\end{array}
	\label{eq:refname24}
\end{equation}

Then evaluating this integral and setting $ {\kappa _0} \to {0^ + } $, $ {\kappa _l} \to \infty $, the wave structure function for a plane wave propagating along the weak ANK horizontal link with anisotropic tilt angle $\gamma$ has the subsequent expression:
\begin{equation}
	D\left( \rho  \right) =  - {2^{3 - \alpha }}\pi {k^2}LA\left( \alpha  \right)\tilde C_n^2\frac{{\Gamma \left( {1 - \frac{\alpha }{2}} \right)}}{{\Gamma \left( {\frac{\alpha }{2}} \right)}}{\rho ^{\alpha  - 2}}\int_0^{2\pi } {{{\left( {\frac{{{{\cos }^2}\theta }}{{\mu _x^2}} + \frac{{{{\sin }^2}\theta }}{{\mu _y^2}}} \right)}^{\frac{\alpha }{2} - 1}}} d\theta.
	\label{eq:refname25}
\end{equation}

On the basis of spatial coherence radius definition, i.e., $ D\left( {{\rho _0}} \right) = 2 $, we can get the spatial coherence radius for a plane wave propagating along the weak ANK horizontal link with anisotropic tilt angle $\gamma$, which has the subsequent expression:
\begin{equation}
	{\rho _0} = {\left[ { - {2^{2 - \alpha }}\pi {k^2}LA\left( \alpha  \right)\tilde C_n^2\frac{{\Gamma \left( {1 - \frac{\alpha }{2}} \right)}}{{\Gamma \left( {\frac{\alpha }{2}} \right)}}\int_0^{2\pi } {{{\left( {\frac{{{{\cos }^2}\theta }}{{\mu _x^2}} + \frac{{{{\sin }^2}\theta }}{{\mu _y^2}}} \right)}^{\frac{\alpha }{2} - 1}}} d\theta } \right]^{\frac{1}{{2 - \alpha }}}}.
	\label{eq:refname26}
\end{equation}

Differing from Eq.~(\ref{eq:refname19}), the wave structure function of non-Kolmogorov turbulence has another expression \cite{Rao1999}:
\begin{equation}
	D\left( \rho  \right) = \frac{{{2^{\alpha  - 1}}{{\left[ {\Gamma \left( {\frac{\alpha }{2} + 1} \right)} \right]}^2}\Gamma \left( {\frac{\alpha }{2} + 2} \right)}}{{\Gamma \left( {\frac{\alpha }{2}} \right)\Gamma \left( {\alpha  + 1} \right)}}{\left( {\frac{\rho }{{{c_0}}}} \right)^{\alpha  - 2}}.
	\label{eq:refname27}
\end{equation}

Then by use of the relationship $D\left( {{\rho _0}} \right) = 2$ and Eq.~(\ref{eq:refname27}), the Fried coherence length $c_0$ has the subsequent expression:
\begin{equation}
	{c_0} = {\left\{ {\frac{{{2^{\alpha  - 2}}{{\left[ {\Gamma \left( {\frac{\alpha }{2} + 1} \right)} \right]}^2}\Gamma \left( {\frac{\alpha }{2} + 2} \right)}}{{\Gamma \left( {\frac{\alpha }{2}} \right)\Gamma \left( {\alpha  + 1} \right)}}} \right\}^{\frac{1}{{\alpha  - 2}}}}{\rho _0}.
	\label{eq:refname28}
\end{equation}
For $ \alpha $ = 11/3, Eq.~(\ref{eq:refname28}) reduces to the form ${c_0} = 2.1{\rho _0}$ which is in line with the Kolmogorov result \cite{Andrews2005}. Finally inserting Eq.~(\ref{eq:refname26}) into Eq.~(\ref{eq:refname28}), we can get the Fried coherence length $c_0$ for a plane wave transmitting along the weak ANK horizontal link with anisotropic tilt angle $\gamma$.

On the basis of the large number theorem, $a_r$ and $a_i$ are asymptotically Gaussian variables, thus the PDF of $a^2$ has the subsequent expression:
\begin{equation}
	\begin{array}{l}
		{f_{{a^2}}}\left( {{a^2}} \right) = \frac{1}{{4\pi {\sigma _r}{\sigma _i}\sqrt {1 - \rho _{{a_r},{a_i}}^2} }}\\
		\hspace*{4.5em}\times \int_0^{2\pi } {\exp \left[ { - \frac{{{{\left( {\frac{{a\cos \theta  - {{\bar a}_r}}}{{{\sigma _r}}}} \right)}^2} + {{\left( {\frac{{a\sin \theta  - {{\bar a}_i}}}{{{\sigma _i}}}} \right)}^2} - 2{\rho _{{a_r},{a_i}}}\left( {\frac{{a\cos \theta  - {{\bar a}_r}}}{{{\sigma _r}}}} \right)\left( {\frac{{a\sin \theta  - {{\bar a}_i}}}{{{\sigma _i}}}} \right)}}{{2\left( {1 - \rho _{{a_r},{a_i}}^2} \right)}}} \right]} d\theta ,
	\end{array}
	\label{eq:refname29}
\end{equation}
where $\sigma _r^2$ and $\sigma _i^2$ represent the variances of $a_r$ and $a_i$, ${\bar a}_r$ and ${\bar a}_i$ represent the means of $a_r$ and $a_i$, ${\rho _{{a_r},{a_i}}} = {\mathop{\rm cov}} \left( {{a_r},{a_i}} \right)/\left( {{\sigma _r}{\sigma _i}} \right)$ represents the correlation coefficient, ${\mathop{\rm cov}} \left( {{a_r},{a_i}} \right)$ represents the covariance of $a_r$ and $a_i$. To estimate these covariance, means, and variances, it can be recalled that $a_r$ and $a_i$ are able to be regarded as the real and imaginary parts for a random phasor. Consequently, we can estimate the covariance, means, and variances for $a_r$ and $a_i$ based on the classical statistical model for speckle with a non-uniform distribution of phases \cite{Goodman2007}. Using the integral formulas \cite{Gradshteyn2000},
\begin{equation}
	\int_0^{2\pi } {\exp \left( { \pm ix\cos \theta } \right)} d\theta  = 2\pi {J_0}\left( x \right),
	\label{eq:refname30}
\end{equation}
\begin{equation}
	\begin{array}{l}
		\int_0^\infty  {x\exp \left( { - \alpha '{x^2}} \right)} {I_\nu }\left( {\beta x} \right){J_\nu }\left( {\gamma x} \right)dx \\
		\hspace*{9em} = \frac{1}{{2\alpha '}}\exp \left( {\frac{{{\beta ^2} - {\gamma ^2}}}{{4\alpha '}}} \right){J_\nu }\left( {\frac{{\gamma \beta }}{{2\alpha '}}} \right),\hspace*{1em}{\mathop{\rm Re}\nolimits} \alpha ' > 0,\hspace*{1em}{\mathop{\rm Re}\nolimits} \nu  >  - 1,
	\end{array}
	\label{eq:refname31}
\end{equation}
and after some algebraic manipulation, the covariance, means, and variances have the subsequent expressions:
\begin{equation}
	{\bar a_r} = \overline {\exp \left[ { - {{\left( {\frac{{\pi {W_m}}}{{\lambda f}}} \right)}^2}r_n^2} \right]} \frac{{8{M_\phi }\left( 1 \right)}}{{{D^2}}}\int_0^{\frac{D}{2}} {\exp \left( { - \frac{{2{\pi ^2}{r^2}\sigma _e^2}}{{{\lambda ^2}{f^2}}}} \right){J_0}\left( {\frac{{2\pi {r_0}r}}{{\lambda f}}} \right)} rdr,
	\label{eq:refname32}
\end{equation}
\begin{equation}
	{\bar a_i} = 0,
	\label{eq:refname33}
\end{equation}
\begin{equation}
	\begin{array}{l}
		\sigma _r^2 = \frac{1}{N}\overline {\exp \left[ { - 2{{\left( {\frac{{\pi {W_m}}}{{\lambda f}}} \right)}^2}r_n^2} \right]} \left[ {\frac{{4{M_\phi }\left( 2 \right)}}{{{D^2}}}\int_0^{\frac{D}{2}} {\exp \left( { - \frac{{8{\pi ^2}{r^2}\sigma _e^2}}{{{\lambda ^2}{f^2}}}} \right){J_0}\left( {\frac{{4\pi {r_0}r}}{{\lambda f}}} \right)} rdr + \frac{1}{2}} \right]\\
		\hspace*{2em}- \frac{1}{N}{\overline {\exp \left[ { - {{\left( {\frac{{\pi {W_m}}}{{\lambda f}}} \right)}^2}r_n^2} \right]} ^2}{\left[ {\frac{{8{M_\phi }\left( 1 \right)}}{{{D^2}}}\int_0^{\frac{D}{2}} {\exp \left( { - \frac{{2{\pi ^2}{r^2}\sigma _e^2}}{{{\lambda ^2}{f^2}}}} \right){J_0}\left( {\frac{{2\pi {r_0}r}}{{\lambda f}}} \right)} rdr} \right]^2},
	\end{array}
	\label{eq:refname34}
\end{equation}
\begin{equation}	
	\sigma _i^2 = \frac{1}{N}\overline {\exp \left[ { - 2{{\left( {\frac{{\pi {W_m}}}{{\lambda f}}} \right)}^2}r_n^2} \right]} \left[ {\frac{1}{2} - \frac{{4{M_\phi }\left( 2 \right)}}{{{D^2}}}\int_0^{\frac{D}{2}} {\exp \left( { - \frac{{8{\pi ^2}{r^2}\sigma _e^2}}{{{\lambda ^2}{f^2}}}} \right){J_0}\left( {\frac{{4\pi {r_0}r}}{{\lambda f}}} \right)} rdr} \right],
	\label{eq:refname35}
\end{equation}
\begin{equation}
	{\mathop{\rm cov}} ({a_r},{a_i}) = 0,
	\label{eq:refname36}
\end{equation}
where ${M_\phi }\left( \omega  \right)$ represents the characteristic function of phase, i.e., the Fourier transform of its PDF, and $\sigma _\phi ^2$ represents the phase variance. Because $a_r$ and $a_i$ are induced by atmospheric turbulence, it can be assumed that phases ${\phi _n}$ follow zero-mean Gaussian statistics, ${p_\phi }\left( \phi  \right) = 1/\left( {\sqrt {2\pi } {\sigma _\phi }} \right)\exp \left[ { - {\phi ^2}/\left( {2\sigma _\phi ^2} \right)} \right]$. Hence, the characteristic function of phase has the subsequent expression \cite{Goodman2007}:
\begin{equation}
	{M_\phi }\left( \omega  \right) = \exp \left( { - \frac{{\sigma _\phi ^2{\omega ^2}}}{2}} \right).
	\label{eq:refname37}
\end{equation}

For the coherent FSO systems, we generally use the phase compensation technique to diminish the phase fluctuations generated by atmospheric turbulence. For non-Kolmogorov turbulence, the residual phase variance after phase compensation can be expressed as \cite{Boreman1996}
\begin{equation}
	\sigma _\phi ^2 = {C_H}{\left( {\frac{D}{{{c_0}}}} \right)^{\alpha  - 2}},
	\label{eq:refname38}
\end{equation}
where $C_H$ represents a parameter dependent on the number of compensation terms $H$.

By use of the Jacobian transformation, the PDF of fiber-coupling efficiency for a plane wave propagating along the weak ANK horizontal link in the presence of bias error, anisotropic tilt angle $\gamma$, and random angular jitter can be written as
\begin{equation}
	f\left( \eta  \right) = \frac{1}{{4\pi {\eta _0}{\sigma _r}{\sigma _i}}}\int_0^{2\pi } {\exp \left[ { - \frac{{{{\left( {{\eta ^{\frac{1}{2}}}\eta _0^{ - \frac{1}{2}}\cos \theta  - {{\bar a}_r}} \right)}^2}}}{{2\sigma _r^2}} - \frac{{\eta \eta _0^{ - 1}{{\sin }^2}\theta }}{{2\sigma _i^2}}} \right]} d\theta.
	\label{eq:refname39}
\end{equation}

For the weak-to-strong fluctuation regime, we can interpret the normalized irradiance for a plane wave transmitting through atmospheric turbulence as the combined result of two independent gamma-distributed random variables, $I = XY$. The $X$ is generated from large-scale atmospheric effects, and the $Y$ is generated from small-scale atmospheric effects. The normalized PDF of Gamma-Gamma distribution has the subsequent expression \cite{Gappmair2017}:
\begin{equation}
	f\left( I \right) = \frac{{2{{\left( {ab} \right)}^{\left( {a + b} \right)/2}}}}{{\Gamma \left( a \right)\Gamma \left( b \right)}}{I^{\left( {a + b} \right)/2 - 1}}{K_{a - b}}\left( {2\sqrt {abI} } \right),\hspace*{1em}I > 0,
	\label{eq:refname40}
\end{equation}
where ${K_{a - b}}\left( x \right)$ represents the modified Bessel function of the order $a - b$ and second kind, $b$ and $a$ represent the parameters of Gamma-Gamma distribution with the small-scale and large-scale scintillations of a plane wave, which have the subsequent expressions:
\begin{equation}
	a = \frac{1}{{\exp \left( {\sigma _{\ln X}^2} \right) - 1}},\hspace*{1em}b = \frac{1}{{\exp \left( {\sigma _{\ln Y}^2} \right) - 1}},
	\label{eq:refname41}
\end{equation}
where $\sigma _{\ln X}^2$ represents the large-scale log-irradiance scintillations, and $\sigma _{\ln Y}^2$ represents the small-scale log-irradiance scintillations. For a plane wave transmitting along the weak-to-strong non-Kolmogorov horizontal link, they have the subsequent expressions \cite{Toselli2007}:
\begin{equation}
	\sigma _{\ln X}^2 = 8{\pi ^2}{k^2}\int_0^L {\int_0^\infty  {\kappa {\Phi _n}\left( {\kappa ,\alpha } \right)} } {G_X}\left( {\kappa ,\alpha } \right)\left[ {1 - \cos \left( {\frac{{{\kappa ^2}z}}{k}} \right)} \right]d\kappa dz,
	\label{eq:refname42}
\end{equation}
\begin{equation}
	\sigma _{\ln Y}^2 = 8{\pi ^2}{k^2}\int_0^L {\int_0^\infty  {\kappa {\Phi _n}\left( {\kappa ,\alpha } \right)} } {G_Y}\left( {\kappa ,\alpha } \right)\left[ {1 - \cos \left( {\frac{{{\kappa ^2}z}}{k}} \right)} \right]d\kappa dz.
	\label{eq:refname43}
\end{equation}
The large-scale and small-scale filter functions demonstrated in Eqs.~(\ref{eq:refname42}) and~(\ref{eq:refname43}) can be determined by
\begin{equation}
	{G_X}\left( {\kappa ,\alpha } \right) = \exp \left( { - \frac{{{\kappa ^2}}}{{\kappa _X^2}}} \right),\hspace*{1em}{G_Y}\left( {\kappa ,\alpha } \right) = \frac{{{\kappa ^\alpha }}}{{{{\left( {{\kappa ^2} + \kappa _Y^2} \right)}^{\frac{\alpha }{2}}}}}.
	\label{eq:refname44}
\end{equation}
The quantities $\kappa _X$ and $\kappa _Y$ represent the cutoff spatial frequencies which remove the mid range scale size effects for the moderate-to-strong fluctuation regime.

Since the low-pass and high-pass spatial frequency cutoffs demonstrated in the filter functions~(\ref{eq:refname44}) are explicitly connected with the correlation width and scattering disk of the transmitting optical wave, we can suppose that at any link distance $L$ into the random medium there exists an effective scattering disk $L/k{l_X}$ and an effective correlation width $l_Y$ related, respectively, to the cutoff wave numbers in accordance with
\begin{equation}
	\frac{L}{{k{l_X}}} = \frac{1}{{{\kappa _X}}} \sim \left\{ {\begin{array}{*{20}{c}}
			{\sqrt {L/k} ,\hspace*{1em}\sigma _R^2 \ll 1,}\\
			{L/k{\rho _0},\hspace*{1em}\sigma _R^2 \gg 1,}
	\end{array}} \right.
	\label{eq:refname45}
\end{equation}
\begin{equation}
	{l_Y} = \frac{1}{{{\kappa _Y}}} \sim \left\{ {\begin{array}{*{20}{c}}
			{\sqrt {L/k} ,\hspace*{1em}\sigma _R^2 \ll 1,}\\
			{{\rho _0},\hspace*{2.3em}\sigma _R^2 \gg 1,}
	\end{array}} \right.
	\label{eq:refname46}
\end{equation}
where $\sigma _R^2$ represents the Rytov variance of a plane wave, and its value is commonly used to define the turbulence scintillation regime. When $\sigma _R^2 \ll 1$, atmospheric turbulence is in the weak fluctuation regime, and for $\sigma _R^2 \gg 1$, atmospheric turbulence is in the moderate to strong fluctuation regime.

The geometrical optics approximation can be utilized in the large-scale log irradiance~(\ref{eq:refname42}) according to
\begin{equation}
	1 - \cos \left( {\frac{{{\kappa ^2}z}}{k}} \right) \cong \frac{1}{2}{\left( {\frac{{{\kappa ^2}z}}{k}} \right)^2},\hspace*{1em}\kappa  \ll {\kappa _X}.
	\label{eq:refname47}
\end{equation}
Hence, by use of the approximation~(\ref{eq:refname47}), the large-scale log-irradiance variance can be reduced to
\begin{equation}
	\sigma _{\ln X}^2 = 4{\pi ^2}{k^2}\int_0^L {\int_0^\infty  {\kappa {\Phi _n}\left( {\kappa ,\alpha } \right)} } {\left( {\frac{{{\kappa ^2}z}}{k}} \right)^2}\exp \left( { - \frac{{{\kappa ^2}}}{{\kappa _X^2}}} \right)d\kappa dz.
	\label{eq:refname48}
\end{equation}

Due to the Markov approximation which supposes that the refraction index is delta-correlated at any pair of points situated along the propagation direction, we can ignore the component $\kappa _z$ in Eq.~(\ref{eq:refname7}). Substituting Eq.~(\ref{eq:refname7}) into Eqs.~(\ref{eq:refname43}) and~(\ref{eq:refname48}) results in
\begin{equation}
	\sigma _{\ln X}^2 = 2\pi {k^2}\int_0^L {\int_0^\infty  {\int_0^\infty  {{\Phi _n}\left( {\kappa ,\alpha ,{\mu _x},{\mu _y}} \right)} } } {\left( {\frac{{{\kappa ^2}z}}{k}} \right)^2}\exp \left( { - \frac{{{\kappa ^2}}}{{\kappa _X^2}}} \right)d{\kappa _x}d{\kappa _y}dz,
	\label{eq:refname49}
\end{equation}
\begin{equation}
	\sigma _{\ln Y}^2 = 4\pi {k^2}\int_0^L {\int_0^\infty  {\int_0^\infty  {{\Phi _n}\left( {\kappa ,\alpha ,{\mu _x},{\mu _y}} \right)} } } \frac{{{\kappa ^\alpha }}}{{{{\left( {{\kappa ^2} + \kappa _Y^2} \right)}^{\frac{\alpha }{2}}}}}\left[ {1 - \cos \left( {\frac{{{\kappa ^2}z}}{k}} \right)} \right]d{\kappa _x}d{\kappa _y}dz.
	\label{eq:refname50}
\end{equation}

We can transform the anisotropic coordinate system of ${\Phi _n}\left( {\kappa ,\alpha ,{\mu _x},{\mu _y}} \right)$ in Eq.~(\ref{eq:refname7}) into an isotropic coordinate system with the help of subsequent mathematical substitutions,
\begin{equation}
	{\kappa _x} = \frac{{{q_x}}}{{{\mu _x}}} = \frac{{q\cos \theta }}{{{\mu _x}}},\hspace*{1em}{\kappa _y} = \frac{{{q_y}}}{{{\mu _y}}} = \frac{{q\sin \theta }}{{{\mu _y}}},\hspace*{1em}q = \sqrt {q_x^2 + q_y^2},
	\label{eq:refname51}
\end{equation}
\begin{equation}
	\kappa  = q\sqrt {\frac{{{{\cos }^2}\theta }}{{\mu _x^2}} + \frac{{{{\sin }^2}\theta }}{{\mu _y^2}}},\hspace*{1em}d{\kappa _x}d{\kappa _y} = \frac{{d{q_x}d{q_y}}}{{{\mu _x}{\mu _y}}} = \frac{{qdqd\theta }}{{{\mu _x}{\mu _y}}},
	\label{eq:refname52}
\end{equation}
\begin{equation}
	{\Phi _n}\left( {\kappa ,\alpha ,{\mu _x},{\mu _y}} \right) = A\left( \alpha  \right)\tilde C_n^2{\mu _x}{\mu _y}{q^{ - \alpha }}.
	\label{eq:refname53}
\end{equation}

Inserting Eqs.~(\ref{eq:refname51}),~(\ref{eq:refname52}), and~(\ref{eq:refname53}) into Eqs.~(\ref{eq:refname49}) and~(\ref{eq:refname50}), we can get
\begin{equation}
	\begin{array}{l}
		\sigma _{\ln X}^2 = 2\pi {k^2}A\left( \alpha  \right)\tilde C_n^2\int_0^L {\int_0^{2\pi } {\int_0^\infty  {{{\left[ {\frac{z}{k}\left( {\frac{{{{\cos }^2}\theta }}{{\mu _x^2}} + \frac{{{{\sin }^2}\theta }}{{\mu _y^2}}} \right)} \right]}^2}} } } \\
		\hspace*{3em}\times {q^{5 - \alpha }}\exp \left[ { - \frac{{{q^2}}}{{\kappa _X^2}}\left( {\frac{{{{\cos }^2}\theta }}{{\mu _x^2}} + \frac{{{{\sin }^2}\theta }}{{\mu _y^2}}} \right)} \right]dqd\theta dz,
	\end{array}
	\label{eq:refname54}
\end{equation}
\begin{equation}
	\begin{array}{l}
		\sigma _{\ln Y}^2 = 4\pi {k^2}A\left( \alpha  \right)\tilde C_n^2\int_0^L {\int_0^{2\pi } {\int_0^\infty  {q{{\left( {\frac{{{{\cos }^2}\theta }}{{\mu _x^2}} + \frac{{{{\sin }^2}\theta }}{{\mu _y^2}}} \right)}^{\frac{\alpha }{2}}}} } } \\
		\hspace*{3em}\times {\left[ {{q^2}\left( {\frac{{{{\cos }^2}\theta }}{{\mu _x^2}} + \frac{{{{\sin }^2}\theta }}{{\mu _y^2}}} \right) + \kappa _Y^2} \right]^{ - \frac{\alpha }{2}}}\left\{ {1 - \cos \left[ {\frac{{{q^2}z}}{k}\left( {\frac{{{{\cos }^2}\theta }}{{\mu _x^2}} + \frac{{{{\sin }^2}\theta }}{{\mu _y^2}}} \right)} \right]} \right\}dqd\theta dz.
	\end{array}
	\label{eq:refname55}
\end{equation}

For the small-scale log-irradiance variance~(\ref{eq:refname55}), we can utilize the approximation \cite{Toselli2007},
\begin{equation}
	\begin{array}{l}
		\int_0^L {\int_0^{2\pi } {\int_0^\infty  {q{{\left( {\frac{{{{\cos }^2}\theta }}{{\mu _x^2}} + \frac{{{{\sin }^2}\theta }}{{\mu _y^2}}} \right)}^{\frac{\alpha }{2}}}} } } {\left[ {{q^2}\left( {\frac{{{{\cos }^2}\theta }}{{\mu _x^2}} + \frac{{{{\sin }^2}\theta }}{{\mu _y^2}}} \right) + \kappa _Y^2} \right]^{ - \frac{\alpha }{2}}}\\
		\times \left\{ {1 - \cos \left[ {\frac{{{q^2}z}}{k}\left( {\frac{{{{\cos }^2}\theta }}{{\mu _x^2}} + \frac{{{{\sin }^2}\theta }}{{\mu _y^2}}} \right)} \right]} \right\}dqd\theta dz\\
		\cong \int_0^L {\int_0^{2\pi } {\int_0^\infty  {q{{\left( {\frac{{{{\cos }^2}\theta }}{{\mu _x^2}} + \frac{{{{\sin }^2}\theta }}{{\mu _y^2}}} \right)}^{\frac{\alpha }{2}}}} } } {\left[ {{q^2}\left( {\frac{{{{\cos }^2}\theta }}{{\mu _x^2}} + \frac{{{{\sin }^2}\theta }}{{\mu _y^2}}} \right) + \kappa _Y^2} \right]^{ - \frac{\alpha }{2}}}dqd\theta dz,\hspace*{1em}{\kappa _Y} \gg \sqrt {\frac{k}{L}}. 
	\end{array}
	\label{eq:refname56}
\end{equation}

Using Eq.~(\ref{eq:refname56}), the small-scale log-irradiance scintillation~(\ref{eq:refname55}) results in
\begin{equation}
	\begin{array}{l}
		\sigma _{\ln Y}^2 = 4\pi {k^2}A\left( \alpha  \right)\tilde C_n^2\int_0^L {\int_0^{2\pi } {\int_0^\infty  {q{{\left( {\frac{{{{\cos }^2}\theta }}{{\mu _x^2}} + \frac{{{{\sin }^2}\theta }}{{\mu _y^2}}} \right)}^{\frac{\alpha }{2}}}} } } \\
		\hspace*{3em}\times {\left[ {{q^2}\left( {\frac{{{{\cos }^2}\theta }}{{\mu _x^2}} + \frac{{{{\sin }^2}\theta }}{{\mu _y^2}}} \right) + \kappa _Y^2} \right]^{ - \frac{\alpha }{2}}}dqd\theta dz.
	\end{array}
	\label{eq:refname57}
\end{equation}

Then using the integral relation \cite{Andrews2005},
\begin{equation}
	\int_0^\infty  {{e^{ - st}}{t^{x - 1}}dt}  = \frac{{\Gamma \left( x \right)}}{{{s^x}}},\hspace*{1em}x > 0,\hspace*{1em}s > 0,
	\label{eq:refname58}
\end{equation}
Eqs.~(\ref{eq:refname54}) and~(\ref{eq:refname57}) can be expressed as
\begin{equation}
	\sigma _{\ln X}^2 = \frac{1}{3}\pi {L^3}A\left( \alpha  \right)\tilde C_n^2\Gamma \left( {3 - \frac{\alpha }{2}} \right)\kappa _X^{6 - \alpha }\int_0^{2\pi } {{{\left( {\frac{{{{\cos }^2}\theta }}{{\mu _x^2}} + \frac{{{{\sin }^2}\theta }}{{\mu _y^2}}} \right)}^{\frac{\alpha }{2} - 1}}} d\theta,
	\label{eq:refname59}
\end{equation}
\begin{equation}
	\sigma _{\ln Y}^2 = \frac{4}{{\alpha  - 2}}\pi {k^2}LA\left( \alpha  \right)\tilde C_n^2\kappa _Y^{2 - \alpha }\int_0^{2\pi } {{{\left( {\frac{{{{\cos }^2}\theta }}{{\mu _x^2}} + \frac{{{{\sin }^2}\theta }}{{\mu _y^2}}} \right)}^{\frac{\alpha }{2} - 1}}} d\theta.
	\label{eq:refname60}
\end{equation}

To evaluate the cutoff wave number $\kappa _X$ and $\kappa _Y$, the asymptotic results~(\ref{eq:refname45}) and~(\ref{eq:refname46}) can be utilized in accordance with
\begin{equation}
	\frac{1}{{\kappa _X^2}} = \frac{{{c_1}L}}{k} + {c_2}{\left( {\frac{L}{{k{\rho _0}}}} \right)^2},
	\label{eq:refname61}
\end{equation}
\begin{equation}
	\kappa _Y^2 = \frac{{{c_3}k}}{L} + \frac{{{c_4}}}{{\rho _0^2}},
	\label{eq:refname62}
\end{equation}
where $c_1$, $c_2$, $c_3$, and $c_4$ represent the scaling constants which can be deduced from the asymptotic behavior given by \cite{Toselli2007}
\begin{equation}
	\sigma _{\ln X}^2 \cong 0.49\sigma _R^2,\hspace*{1em}\sigma _{\ln Y}^2 \cong 0.51\sigma _R^2,\hspace*{1em}\sigma _R^2 \ll 1,
	\label{eq:refname63}
\end{equation}
\begin{equation}
	\sigma _{\ln X}^2 \cong 0.5\sigma _I^2 - 0.5,\hspace*{1em}\sigma _{\ln Y}^2 \cong \ln 2,\hspace*{1em}\sigma _R^2 \gg 1,
	\label{eq:refname64}
\end{equation}
where $\sigma _I^2$ represents the scintillation index of a plane wave. Notice that, from Ref. \cite{Andrews2005}, we can obtain the well-known result that the scintillation index of a plane wave in the weak fluctuation regime is equal to the Rytov variance. Then the Rytov variance of a plane wave with anisotropic tilt angle $\gamma$ in the ANK horizontal link has the subsequent expression \cite{Zhai2021b}:
\begin{equation}
	\begin{array}{l}
		\sigma _R^2 =  - \frac{2}{\alpha }\Gamma \left( {1 - \frac{\alpha }{2}} \right)\Gamma \left( {\alpha  - 1} \right)\sin \left( {\frac{{\alpha \pi }}{4}} \right)\cos \left( {\frac{{\alpha \pi }}{2}} \right)\tilde C_n^2{k^{3 - \frac{\alpha }{2}}}{L^{\frac{\alpha }{2}}}\\
		\hspace*{2.3em}\times \frac{1}{{2\pi }}\int_0^{2\pi } {{{\left( {\frac{{{{\cos }^2}\theta }}{{\mu _x^2}} + \frac{{{{\sin }^2}\theta }}{{\mu _y^2}}} \right)}^{\frac{\alpha }{2} - 1}}} d\theta.
	\end{array}
	\label{eq:refname65}
\end{equation}

Substituting Eqs.~(\ref{eq:refname59}),~(\ref{eq:refname60}),~(\ref{eq:refname61}),~(\ref{eq:refname62}), and~(\ref{eq:refname65}) into Eqs.~(\ref{eq:refname63}) and~(\ref{eq:refname64}), we can obtain
\begin{equation}
	{c_1} = {\left[ { - \frac{{5.88}}{\alpha }\frac{{\Gamma \left( {1 - \frac{\alpha }{2}} \right)}}{{\Gamma \left( {3 - \frac{\alpha }{2}} \right)}}\sin \left( {\frac{{\alpha \pi }}{4}} \right)} \right]^{\frac{2}{{\alpha  - 6}}}},
	\label{eq:refname66}
\end{equation}
\begin{equation}
	{c_3} = {\left[ { - \frac{{0.51\left( {\alpha  - 2} \right)}}{\alpha }\Gamma \left( {1 - \frac{\alpha }{2}} \right)\sin \left( {\frac{{\alpha \pi }}{4}} \right)} \right]^{\frac{2}{{2 - \alpha }}}},
	\label{eq:refname67}
\end{equation}
\begin{equation}
	{c_4} = {\left[ {\frac{{\left( {2 - \alpha } \right)\Gamma \left( {1 - \frac{\alpha }{2}} \right)}}{{{2^\alpha }\Gamma \left( {\frac{\alpha }{2}} \right)}}\ln 2} \right]^{\frac{2}{{2 - \alpha }}}}.
	\label{eq:refname68}
\end{equation}

With the aim of calculating the scaling constant $c_2$, the scintillation index of a plane wave $\sigma _I^2$ for $\sigma _R^2 \gg 1$ is necessary. In the saturation regime that is for the high value of Rytov variance, the scintillation index of a plane wave has the subsequent expression \cite{Andrews2005}:
\begin{equation}
	\begin{array}{l}
		\sigma _I^2 = 1 + 32{\pi ^2}{k^2}L\int_0^1 {\int_0^\infty  {\kappa {\Phi _n}} } \left( \kappa  \right){\sin ^2}\left[ {\frac{{L{\kappa ^2}}}{{2k}}w\left( {\xi ,\xi } \right)} \right]\\
		\hspace*{8em}\times \exp \left\{ { - \int_0^1 {{D_S}\left[ {\frac{{L\kappa }}{k}w\left( {\varsigma ,\xi } \right)} \right]} d\varsigma } \right\}d\kappa d\xi ,\hspace*{1em}\sigma _R^2 \gg 1,
	\end{array}
	\label{eq:refname69}
\end{equation}
where $\varsigma$ represents a normalized distance variable, and the exponential function behaves as a low-pass spatial filter determined by the plane wave structure function of phase ${D_S}\left( \rho  \right)$. The function $w\left( {\varsigma ,\xi } \right)$ can be expressed as
\begin{equation}
	w\left( {\varsigma ,\xi } \right) = \left\{ {\begin{array}{*{20}{c}}
			{\varsigma ,\hspace*{1em}\varsigma  < \xi ,}\\
			{\xi ,\hspace*{1em}\varsigma  > \xi .}
	\end{array}} \right.
	\label{eq:refname70}
\end{equation}
Notice that, under the geometrical optics approximation $L{\kappa ^2}/k \ll 1$ and for a plane wave, the phase structure function ${D_S}\left( \rho  \right)$ is equivalent to the wave structure function $D\left( \rho  \right)$. Utilizing the supposition that
the spatial coherence radius of a plane wave is larger than the inner scale of turbulence, and on the basis of the wave structure function for a plane wave propagating along the weak ANK horizontal link with anisotropic tilt angle $\gamma$ Eq.~(\ref{eq:refname25}), it follows that
\begin{equation}
	\begin{array}{l}
		\int_0^1 {{D_S}\left[ {\frac{{L\kappa }}{k}w\left( {\varsigma ,\xi } \right)} \right]} d\varsigma  =  - {2^{3 - \alpha }}\pi {k^{4 - \alpha }}{L^{\alpha  - 1}}A\left( \alpha  \right)\tilde C_n^2\frac{{\Gamma \left( {1 - \frac{\alpha }{2}} \right)}}{{\Gamma \left( {\frac{\alpha }{2}} \right)}}\\
		\hspace*{10.5em}\times {\kappa ^{\alpha  - 2}}{\xi ^{\alpha  - 2}}\left( {1 + \frac{{2 - \alpha }}{{\alpha  - 1}}\xi } \right)\int_0^{2\pi } {{{\left( {\frac{{{{\cos }^2}\theta }}{{\mu _x^2}} + \frac{{{{\sin }^2}\theta }}{{\mu _y^2}}} \right)}^{\frac{\alpha }{2} - 1}}d\theta } .
	\end{array}
	\label{eq:refname71}
\end{equation}
Similarly, we may approximate the sine function in Eq.~(\ref{eq:refname69}) by its leading term, which leads to
\begin{equation}
	{\sin ^2}\left( {\frac{{L{\kappa ^2}\xi }}{{2k}}} \right) \cong \frac{{{L^2}{\kappa ^4}{\xi ^2}}}{{4{k^2}}}.
	\label{eq:refname72}
\end{equation}
Inserting Eqs.~(\ref{eq:refname71}) and~(\ref{eq:refname72}) into Eq.~(\ref{eq:refname69}), we can obtain
\begin{equation}
	\begin{array}{l}
		\sigma _I^2 = 1 + 8{\pi ^2}{L^3}\int_0^1 {\int_0^\infty  {{\kappa ^5}{\Phi _n}} } \left( \kappa  \right){\xi ^2}\exp \left[ {{2^{3 - \alpha }}\pi {k^{4 - \alpha }}{L^{\alpha  - 1}}A\left( \alpha  \right)\tilde C_n^2\frac{{\Gamma \left( {1 - \frac{\alpha }{2}} \right)}}{{\Gamma \left( {\frac{\alpha }{2}} \right)}}} \right.\\
		\hspace*{2em}\left. { \times {\kappa ^{\alpha  - 2}}{\xi ^{\alpha  - 2}}\left( {1 + \frac{{2 - \alpha }}{{\alpha  - 1}}\xi } \right)\int_0^{2\pi } {{{\left( {\frac{{{{\cos }^2}\theta }}{{\mu _x^2}} + \frac{{{{\sin }^2}\theta }}{{\mu _y^2}}} \right)}^{\frac{\alpha }{2} - 1}}d\theta } } \right]d\kappa d\xi ,\hspace*{1em}\sigma _R^2 \gg 1.
	\end{array}
	\label{eq:refname73}
\end{equation}

Due to the Markov approximation which supposes that the refraction index is delta-correlated at any pair of points situated along the propagation direction, we can ignore the component $\kappa _z$ in Eq.~(\ref{eq:refname7}). Substituting Eq.~(\ref{eq:refname7}) into Eq.~(\ref{eq:refname73}) results in
\begin{equation}
	\begin{array}{l}
		\sigma _I^2 = 1 + 4\pi {L^3}\int_0^1 {\int_0^\infty  {\int_0^\infty  {{\kappa ^4}} {\Phi _n}\left( {\kappa ,\alpha ,{\mu _x},{\mu _y}} \right)} } {\xi ^2}\exp \left[ {{2^{3 - \alpha }}\pi {k^{4 - \alpha }}{L^{\alpha  - 1}}A\left( \alpha  \right)\tilde C_n^2} \right.\\
		\hspace*{1.5em}\left. { \times \frac{{\Gamma \left( {1 - \frac{\alpha }{2}} \right)}}{{\Gamma \left( {\frac{\alpha }{2}} \right)}}{\kappa ^{\alpha  - 2}}{\xi ^{\alpha  - 2}}\left( {1 + \frac{{2 - \alpha }}{{\alpha  - 1}}\xi } \right)\int_0^{2\pi } {{{\left( {\frac{{{{\cos }^2}\theta }}{{\mu _x^2}} + \frac{{{{\sin }^2}\theta }}{{\mu _y^2}}} \right)}^{\frac{\alpha }{2} - 1}}d\theta } } \right]d{\kappa _x}d{\kappa _y}d\xi ,\hspace*{0.5em}\sigma _R^2 \gg 1.
	\end{array}
	\label{eq:refname74}
\end{equation}

We can transform the anisotropic coordinate system of ${\Phi _n}\left( {\kappa ,\alpha ,{\mu _x},{\mu _y}} \right)$ in Eq.~(\ref{eq:refname7}) into an isotropic coordinate system with the help of mathematical substitutions~(\ref{eq:refname51}),~(\ref{eq:refname52}), and~(\ref{eq:refname53}). Substituting Eqs.~(\ref{eq:refname51}),~(\ref{eq:refname52}), and~(\ref{eq:refname53}) into Eq.~(\ref{eq:refname74}), we can obtain
\begin{equation}
	\begin{array}{l}
		\sigma _I^2 = 1 + 4\pi {L^3}A\left( \alpha  \right)\tilde C_n^2\int_0^1 {\int_0^{2\pi } {\int_0^\infty  {{q^{5 - \alpha }}{{\left( {\frac{{{{\cos }^2}\theta }}{{\mu _x^2}} + \frac{{{{\sin }^2}\theta }}{{\mu _y^2}}} \right)}^2}} } } {\xi ^2}\\
		\hspace*{2.2em}\times \exp \left[ {{2^{3 - \alpha }}\pi {k^{4 - \alpha }}{L^{\alpha  - 1}}A\left( \alpha  \right)\tilde C_n^2\frac{{\Gamma \left( {1 - \frac{\alpha }{2}} \right)}}{{\Gamma \left( {\frac{\alpha }{2}} \right)}}{q^{\alpha  - 2}}{{\left( {\frac{{{{\cos }^2}\theta }}{{\mu _x^2}} + \frac{{{{\sin }^2}\theta }}{{\mu _y^2}}} \right)}^{\frac{\alpha }{2} - 1}}} \right.\\
		\hspace*{2.2em}\left. { \times {\xi ^{\alpha  - 2}}\left( {1 + \frac{{2 - \alpha }}{{\alpha  - 1}}\xi } \right)\int_0^{2\pi } {{{\left( {\frac{{{{\cos }^2}\theta }}{{\mu _x^2}} + \frac{{{{\sin }^2}\theta }}{{\mu _y^2}}} \right)}^{\frac{\alpha }{2} - 1}}d\theta } } \right]dqd\theta d\xi ,\hspace*{1em}\sigma _R^2 \gg 1.
	\end{array}
	\label{eq:refname75}
\end{equation}

Using the integral formulas Eq.~(\ref{eq:refname58}) and,
\begin{equation}
	\int_0^x {\frac{{{t^{\mu  - 1}}}}{{{{\left( {1 + \beta t} \right)}^v}}}} dt = \frac{{{x^\mu }}}{\mu }{}_2{F_1}\left( {v,\mu ;1 + \mu ; - \beta x} \right),\hspace*{1em}\mu  > 0,
	\label{eq:refname76}
\end{equation}
the scintillation index for a plane wave propagating along the ANK horizontal link with anisotropic tilt angle $\gamma$ in the saturation regime has the subsequent expression:
\begin{equation}
	\begin{array}{l}
		\sigma _I^2 = 1 + {2^{3 - }}^{\frac{4}{{\alpha  - 2}}}\pi {k^{6 - \alpha }}{L^{\alpha  - 3}}A\left( \alpha  \right)\tilde C_n^2\rho _0^{6 - \alpha }\int_0^{2\pi } {{{\left( {\frac{{{{\cos }^2}\theta }}{{\mu _x^2}} + \frac{{{{\sin }^2}\theta }}{{\mu _y^2}}} \right)}^{\frac{\alpha }{2} - 1}}} d\theta \\
		\hspace*{2.2em}\times \frac{{\Gamma \left( {\frac{{6 - \alpha }}{{\alpha  - 2}}} \right){}_2{F_1}\left( {\frac{{6 - \alpha }}{{\alpha  - 2}},\alpha  - 3;\alpha  - 2;\frac{{\alpha  - 2}}{{\alpha  - 1}}} \right)}}{{\left( {\alpha  - 2} \right)\left( {\alpha  - 3} \right)}},\hspace*{1em}\sigma _R^2 \gg 1.
	\end{array}
	\label{eq:refname77}
\end{equation}

Substituting Eqs.~(\ref{eq:refname59}),~(\ref{eq:refname61}), and~(\ref{eq:refname77}) into Eq.~(\ref{eq:refname64}), we can obtain
\begin{equation}
	{c_2} = {\left[ {\frac{{3 \times {2^{2 - }}^{\frac{4}{{\alpha  - 2}}}\Gamma \left( {\frac{{6 - \alpha }}{{\alpha  - 2}}} \right){}_2{F_1}\left( {\frac{{6 - \alpha }}{{\alpha  - 2}},\alpha  - 3;\alpha  - 2;\frac{{\alpha  - 2}}{{\alpha  - 1}}} \right)}}{{\left( {\alpha  - 2} \right)\left( {\alpha  - 3} \right)\Gamma \left( {3 - \frac{\alpha }{2}} \right)}}} \right]^{\frac{2}{{\alpha  - 6}}}}.
	\label{eq:refname78}
\end{equation}

Inserting Eqs.~(\ref{eq:refname61}),~(\ref{eq:refname66}), and~(\ref{eq:refname78}) into Eq.~(\ref{eq:refname59}), the large-scale log-irradiance variance for a plane wave transmitting along the weak-to-strong ANK horizontal link with anisotropic tilt angle $\gamma$ has the subsequent expression:
\begin{equation}
	\sigma _{\ln X}^2 = \frac{{0.49\sigma _R^2}}{{{{\left[ {1 + M\left( \alpha  \right)\sigma _R^{\frac{4}{{\alpha  - 2}}}} \right]}^{3 - \frac{\alpha }{2}}}}},
	\label{eq:refname79}
\end{equation}
where
\begin{equation}
	M\left( \alpha  \right) = {\left\{ {\frac{{1.02\Gamma \left( {\frac{{6 - \alpha }}{{\alpha  - 2}}} \right){}_2{F_1}\left( {\frac{{6 - \alpha }}{{\alpha  - 2}},\alpha  - 3;\alpha  - 2;\frac{{\alpha  - 2}}{{\alpha  - 1}}} \right)}}{{ - \Gamma \left( {1 - \frac{\alpha }{2}} \right)\left( {\alpha  - 2} \right)\left( {\alpha  - 3} \right)}}{{\left[ {\frac{\alpha }{{\sin \left( {\frac{{\alpha \pi }}{4}} \right)}}} \right]}^{\frac{{2\alpha  - 8}}{{\alpha  - 2}}}}\frac{{{2^{5 - \alpha  + \frac{4}{{\alpha  - 2}}}}}}{{{{\left[ {\Gamma \left( {\frac{\alpha }{2}} \right)} \right]}^{\frac{{\alpha  - 6}}{{\alpha  - 2}}}}}}} \right\}^{\frac{2}{{\alpha  - 6}}}}.
	\label{eq:refname80}
\end{equation}

Inserting Eqs.~(\ref{eq:refname62}),~(\ref{eq:refname67}), and~(\ref{eq:refname68}) into Eq.~(\ref{eq:refname60}), the small-scale log-irradiance variance for a plane wave transmitting along the weak-to-strong ANK horizontal link with anisotropic tilt angle $\gamma$ has the subsequent expression:
\begin{equation}
	\sigma _{\ln Y}^2 = \frac{{0.51\sigma _R^2}}{{{{\left[ {1 + {{\left( {1.3591} \right)}^{\frac{2}{{2 - \alpha }}}}\sigma _R^{\frac{4}{{\alpha  - 2}}}} \right]}^{\frac{\alpha }{2} - 1}}}}.
	\label{eq:refname81}
\end{equation}
Notice that, inserting Eqs.~(\ref{eq:refname41}),~(\ref{eq:refname79}), and~(\ref{eq:refname81}) into Eq.~(\ref{eq:refname40}), the normalized PDF of Gamma-Gamma distribution for a plane wave propagating along the weak-to-strong ANK horizontal link with anisotropic tilt angle $\gamma$ is gotten.

\section{Average bit error rate}
The arbitrary $J \times Q$ rectangular QAM is made up of two separate pulse amplitude modulation (PAM) signals. A bit error in a $J \times Q$ rectangular QAM symbol is completely generated by a bit error in either $J$-ary PAM or $Q$-ary PAM. Hence, by averaging the BER of two PAM signals, we can get the average BER of $J \times Q$ rectangular QAM, which can be expressed as \cite{Cho2002}
\begin{equation}
	{P_b} = \frac{1}{{{{\log }_2}\left( {J \cdot Q} \right)}}\left( {\sum\limits_{k = 1}^{{{\log }_2}J} {{P_J}\left( k \right) + \sum\limits_{l = 1}^{{{\log }_2}Q} {{P_Q}\left( l \right)} } } \right),
	\label{eq:refname82}
\end{equation}
where ${{P_J}\left( k \right)}$ and ${{P_Q}\left( l \right)}$ represent severally the error probability for the $k$th bit of in-phase signals and the $l$th bit of quadrature signals.

On the basis of the research about QAM signal transmission through the AWGN channel in \cite{Cho2002}, the error probability for the $k$th bit of in-phase signals and the $l$th bit of quadrature signals can be obtained by use of the SNR per bit $\gamma _b$, which can be expressed as
\begin{equation}
	\begin{array}{l}
		{P_J}\left( k \right) = \frac{1}{J}\sum\limits_{j = 0}^{\left( {1 - {2^{ - k}}} \right)J - 1} {\left\{ {{{\left( { - 1} \right)}^{\left\lfloor {\frac{{j \cdot {2^{k - 1}}}}{J}} \right\rfloor }}\left[ {{2^{k - 1}} - \left\lfloor {\frac{{j \cdot {2^{k - 1}}}}{J} + \frac{1}{2}} \right\rfloor } \right]} \right.} \\
		\hspace*{4em}\left. { \times erfc\left( {\left( {2j + 1} \right)\sqrt {\frac{{3{{\log }_2}\left( {J \cdot Q} \right){\gamma _b}}}{{{J^2} + {Q^2} - 2}}} } \right)} \right\},
	\end{array}
	\label{eq:refname83}
\end{equation}
\begin{equation}
	\begin{array}{l}
		{P_Q}\left( l \right) = \frac{1}{Q}\sum\limits_{q = 0}^{\left( {1 - {2^{ - l}}} \right)Q - 1} {\left\{ {{{\left( { - 1} \right)}^{\left\lfloor {\frac{{q \cdot {2^{l - 1}}}}{Q}} \right\rfloor }}\left[ {{2^{l - 1}} - \left\lfloor {\frac{{q \cdot {2^{l - 1}}}}{Q} + \frac{1}{2}} \right\rfloor } \right]} \right.} \\
		\hspace*{4em}\left. { \times erfc\left( {\left( {2q + 1} \right)\sqrt {\frac{{3{{\log }_2}\left( {J \cdot Q} \right){\gamma _b}}}{{{J^2} + {Q^2} - 2}}} } \right)} \right\},
	\end{array}
	\label{eq:refname84}
\end{equation}
where $\left\lfloor x \right\rfloor$ represents the largest integer of $x$, and $erfc\left( x \right)$ represents the complementary error function. Notice that the signal constellation is arranged on the basis of a perfect two-dimensional Gray code \cite{Weber1978}.

When the value of SNR is high, it can be easily obtained that the terms with $j = 0$ and $q = 0$ in Eqs.~(\ref{eq:refname83}) and~(\ref{eq:refname84}) will be predominant, so we can get an approximate BER expression for $J \times Q$ rectangular QAM from Eq.~(\ref{eq:refname82}) by neglecting the higher order terms, which can be expressed as
\begin{equation}
	\begin{array}{l}
		{P_b} = \frac{1}{{{{\log }_2}\left( {J \cdot Q} \right)}}\left[ {\frac{{J - 1}}{J}erfc\left( {\sqrt {\frac{{3{{\log }_2}\left( {J \cdot Q} \right){\gamma _b}}}{{{J^2} + {Q^2} - 2}}} } \right)} \right.\\
		\hspace*{15em}\left. { + \frac{{Q - 1}}{Q}erfc\left( {\sqrt {\frac{{3{{\log }_2}\left( {J \cdot Q} \right){\gamma _b}}}{{{J^2} + {Q^2} - 2}}} } \right)} \right].
	\end{array}
	\label{eq:refname85}
\end{equation}

Since the fiber-coupling efficiency $\eta$ and the atmospheric turbulence-induced fading $I$ are random, the value of $\gamma _b$ is also random. Then inserting Eqs.~(\ref{eq:refname5}),~(\ref{eq:refname39}), and~(\ref{eq:refname40}) into Eq.~(\ref{eq:refname85}), the approximate average BER expression of fiber-based $J \times Q$ rectangular QAM/FSO systems for a plane wave propagating along the weak ANK horizontal link in the presence of bias error, anisotropic tilt angle $\gamma$, and random angular jitter is developed, which has the subsequent expression:
\begin{equation}
	\begin{array}{l}
		{P_b} = \frac{{2{{\left( {ab} \right)}^{\left( {a + b} \right)/2}}}}{{4\pi {\eta _0}{\sigma _r}{\sigma _i}\Gamma \left( a \right)\Gamma \left( b \right){{\log }_2}\left( {J \cdot Q} \right)}}\left( {\frac{{J - 1}}{J} + \frac{{Q - 1}}{Q}} \right)\int_0^\infty  {\int_0^{0.8145} {\int_0^{2\pi } {{K_{a - b}}\left( {2\sqrt {abI} } \right)} } } \\
		\hspace*{2em}\times {I^{\frac{{a + b}}{2} - 1}}erfc\left[ {\sqrt {\frac{{1.5{R^2}{\eta ^2}{I^2}P_r^2}}{{\sigma _n^2{R_s}\left( {{J^2} + {Q^2} - 2} \right)}}} } \right]\exp \left[ { - \frac{{{{\left( {{\eta ^{\frac{1}{2}}}\eta _0^{ - \frac{1}{2}}\cos \theta  - {{\bar a}_r}} \right)}^2}}}{{2\sigma _r^2}} - \frac{{\eta \eta _0^{ - 1}{{\sin }^2}\theta }}{{2\sigma _i^2}}} \right]d\theta d\eta dI.
	\end{array}
	\label{eq:refname86}
\end{equation}
Notice that, to simplify the analysis, the numerical evaluation in this paper is based upon the approximate average BER expression Eq.~(\ref{eq:refname86}), and the choosing of atmospheric turbulence parameters and communication system parameters guarantees a high SNR value.

\section{Numerical results}
On the basis of the approximate average BER expression for a plane wave propagating through the weak ANK horizontal link in the presence of bias error, anisotropic tilt angle $\gamma$, and random angular jitter Eq.~(\ref{eq:refname86}), we numerically investigate the impacts of atmospheric turbulence parameters and communication system parameters on the BER of fiber-based $J \times Q$ rectangular QAM/FSO systems. Notice that, in the subsequent evaluations, unless the parameters are explicitly exhibited in the figures, we have $ L $ = 1 km, $ \lambda $ = 1.55 $ \mu $m, $ \alpha $ = 3.5, $ \mu $ = 2, $\tilde C_n^2 = 1 \times {10^{ - 14}}{{\rm m}^{3 - \alpha }}$, $ D $ = 0.1 m, $ f $ = 0.4 m, $ W_m $ = 5 $ \mu $m, $ \gamma $ = 45 $ ^\circ $, $ \omega $ = 60 $ ^\circ $, $ \sigma _e $ = 0 $ \mu $m, $ r_0 $ = 0 $ \mu $m, $J$ = 4, $Q$ = 2, $R$ = 0.9 A/W, $P_t$ = 1 W, $ \vartheta $ = 1 mrad, $\sigma _n = 1 \times 10^{ - 9}{\rm A}/{\rm Hz}$, $T_s$ = 1 ns, $ H $ = 3, and $a_l$ = 1 (without energy loss).

\begin{figure}[ht!]
	\centering
	\includegraphics[width=6.6cm]{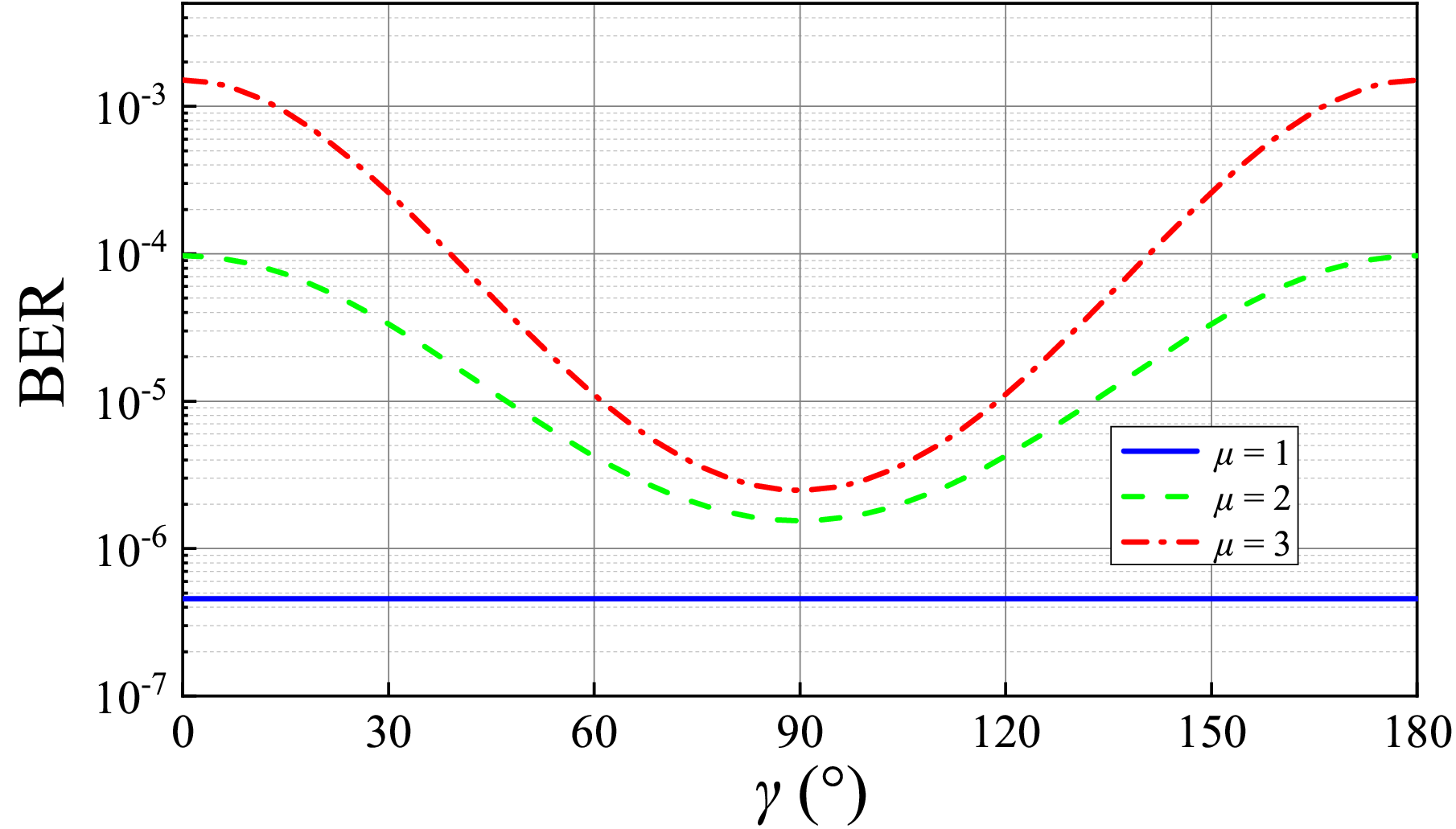}
	\includegraphics[width=6.6cm]{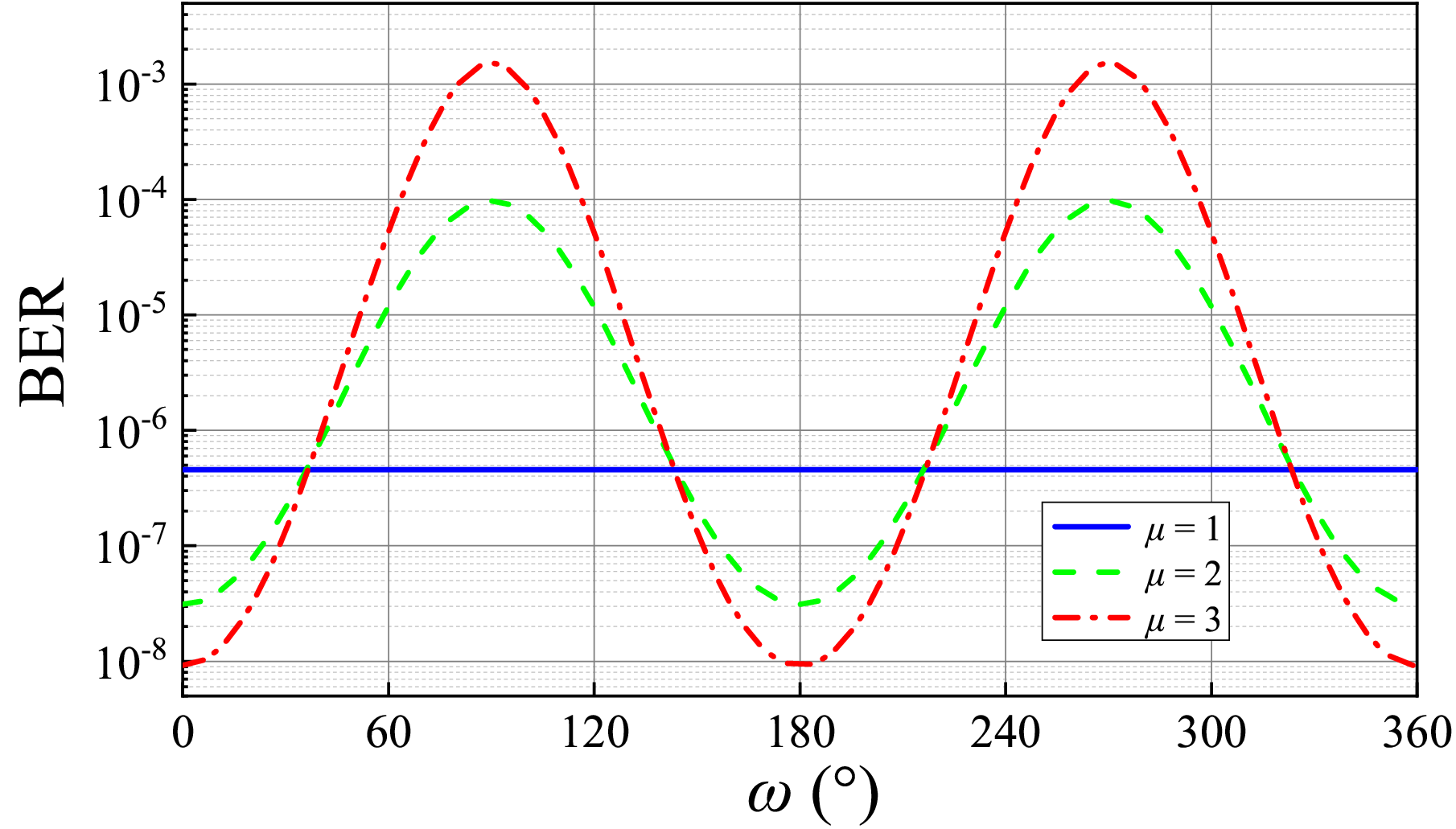}
	\caption{BER of fiber-based QAM/FSO systems under several values of anisotropic factor $\mu$ in the horizontal link. (a) as the function of anisotropic tilt angle $\gamma$; (b) as the function of azimuth angle $\omega$.}
	\label{Fig03}
\end{figure}

Figure~\ref{Fig03} illustrates the average BER of fiber-based QAM/FSO systems in the horizontal link reliance on the anisotropic tilt angle $\gamma$ or the azimuth angle $\omega$ with different anisotropic factor $\mu$ values. We can observe from Fig.~\ref{Fig03} that, while the anisotropic factor $\mu$ is equal to 1, the anisotropic atmospheric turbulence reduces to the isotropic atmospheric turbulence, consequently the changes of azimuth angle $\omega$ and anisotropic tilt angle $\gamma$ have no impact on the average BER of fiber-based QAM/FSO systems, then for the horizontal link, our turbulence spectrum models are self-consistent. For the anisotropic factor $\mu$ larger than 1 for any azimuth angle $\omega$ except $ \omega $ = 90 $ ^\circ $ and 270 $ ^\circ $, the average BER of fiber-based QAM/FSO systems first decays and then increases with the growing of anisotropic tilt angle $\gamma$, and the variation trends of BER are symmetrical about the straight-line $\gamma$ = 90 $ ^\circ $. Similarly, for the anisotropic factor $\mu$ larger than 1 for any anisotropic tilt angle $\gamma$ except $ \gamma $ = 0 $ ^\circ $ and 180 $ ^\circ $, as the azimuth angle $\omega$ increases from 0 $ ^\circ $ to 180 $ ^\circ $, the average BER of fiber-based QAM/FSO systems increases up to a peak value at $\omega$ = 90 $ ^\circ $, and after the peak point, it begins to decay. And the variation trends of BER are symmetrical about the straight-line $\omega$ = 180 $ ^\circ $. We can also obtain from Fig.~\ref{Fig03} that, for $\omega$ = 60 $ ^\circ $ for any value of $\gamma$ (the region is closer to the plane where the long axes of turbulence cell ellipsoid model are located), the average BER of fiber-based QAM/FSO systems increases with the growing of $\mu$. Then for $\gamma$ = 45 $ ^\circ $ for $\omega$ = 0 $ ^\circ $ to 37 $ ^\circ $, 143 $ ^\circ $ to 217 $ ^\circ $, 323 $ ^\circ $ to 360 $ ^\circ $ (the region is closer to the short axis of turbulence cell ellipsoid model), the average BER of fiber-based QAM/FSO systems decays with the growing of $\mu$. But for $\gamma$ = 45 $ ^\circ $ for the rest of $\omega$ range (the region is closer to the plane where the long axes of turbulence cell ellipsoid model are located), the contrary trend will occur.

The physical interpretation for these comments can be interpreted by the fact that the curvature for anisotropic turbulence cells is not similar to the isotropic situation, and the anisotropic turbulence cells will cause changes in the turbulence focusing characteristics. While a beam transmits along the long axis of anisotropic turbulence cells, the radius of curvature for the interface between the anisotropic turbulence cells and the beam is small, and the beam will be more deviated from the propagation direction by reason that these turbulence cells perform as lenses with a lower radius of curvature \cite{Zhai2021b,Toselli2011}. Accordingly, when a beam transmits along the short axis of anisotropic turbulence cells, the radius of curvature for the interface between the anisotropic turbulence cells and the beam is large, and the beam will be less deviated from the propagation direction by reason that these turbulence cells perform as lenses with a higher radius of curvature. Subsequently, on the basis of simple spatial geometry, it can be easily deduced that for $ \gamma $ = 0 $ ^\circ $ and 180 $ ^\circ $, the radius of curvature for the interface between the anisotropic turbulence cells and the beam is basically unchanged with the change of $\omega$ in the horizontal link. Therefore, the average BER of fiber-based QAM/FSO systems remains unchanged. In the same way, for $ \omega $ = 90 $ ^\circ $ and 270 $ ^\circ $, the radius of curvature for the interface between the anisotropic turbulence cells and the beam is also unchanged with the change of $\gamma$ in the horizontal link, which will cause the average BER of fiber-based QAM/FSO systems to remain unchanged. Furthermore, the reason why the average BER of fiber-based QAM/FSO systems varies periodically with $\gamma$ and $\omega$ can be simply explained by the symmetry of turbulence cell ellipsoid model combined with the above physical theory. In addition, the intersections of the average BER curves under different anisotropic factors $\mu$ are not fixed at the azimuth angle $\omega$ = 37 $ ^\circ $, 143 $ ^\circ $, 217 $ ^\circ $, and 323 $ ^\circ $, which will change with the variation of anisotropic tilt angle $\gamma$.

\begin{figure}[ht!]
	\centering
	\includegraphics[width=6.6cm]{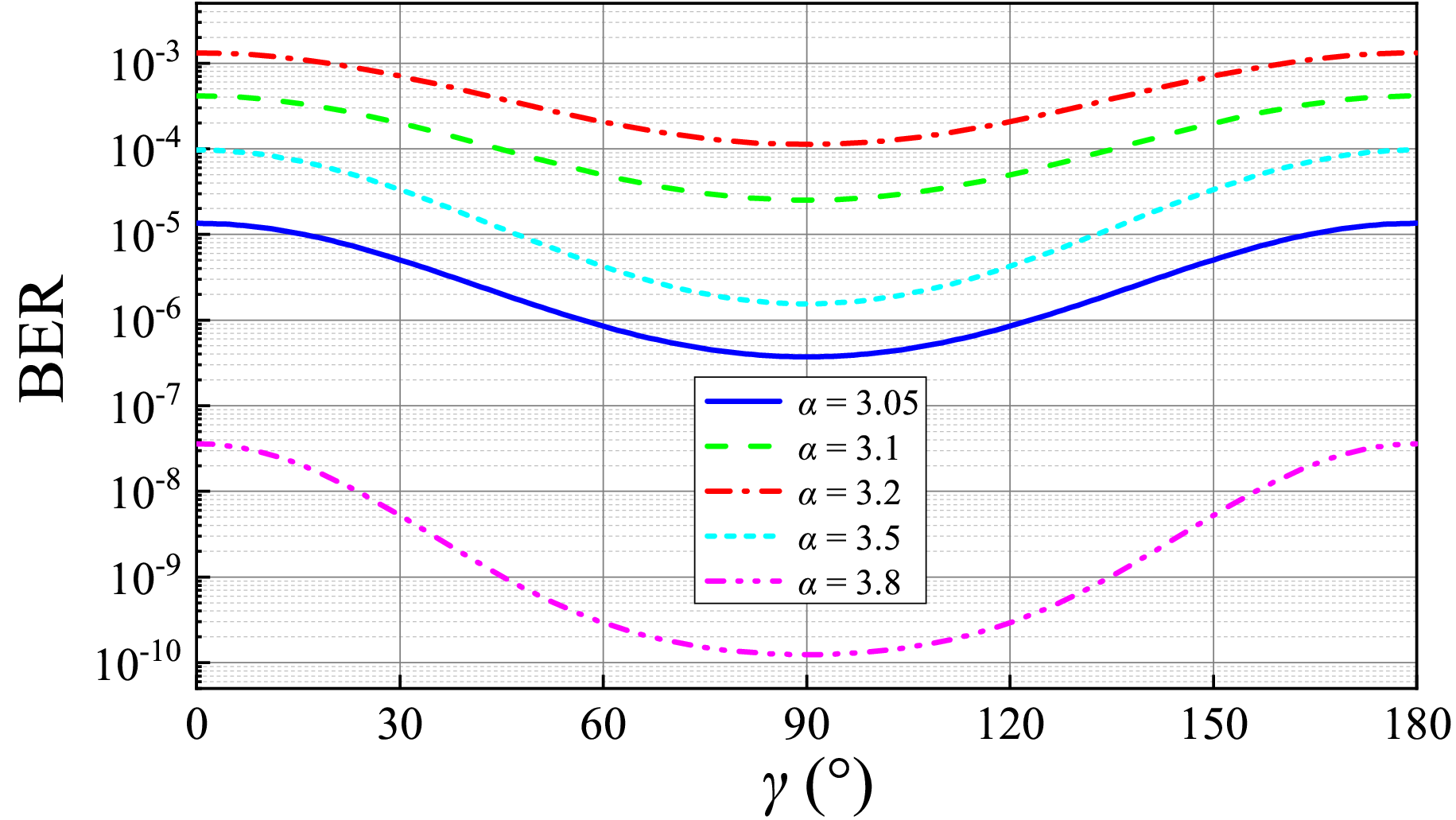}
	\includegraphics[width=6.6cm]{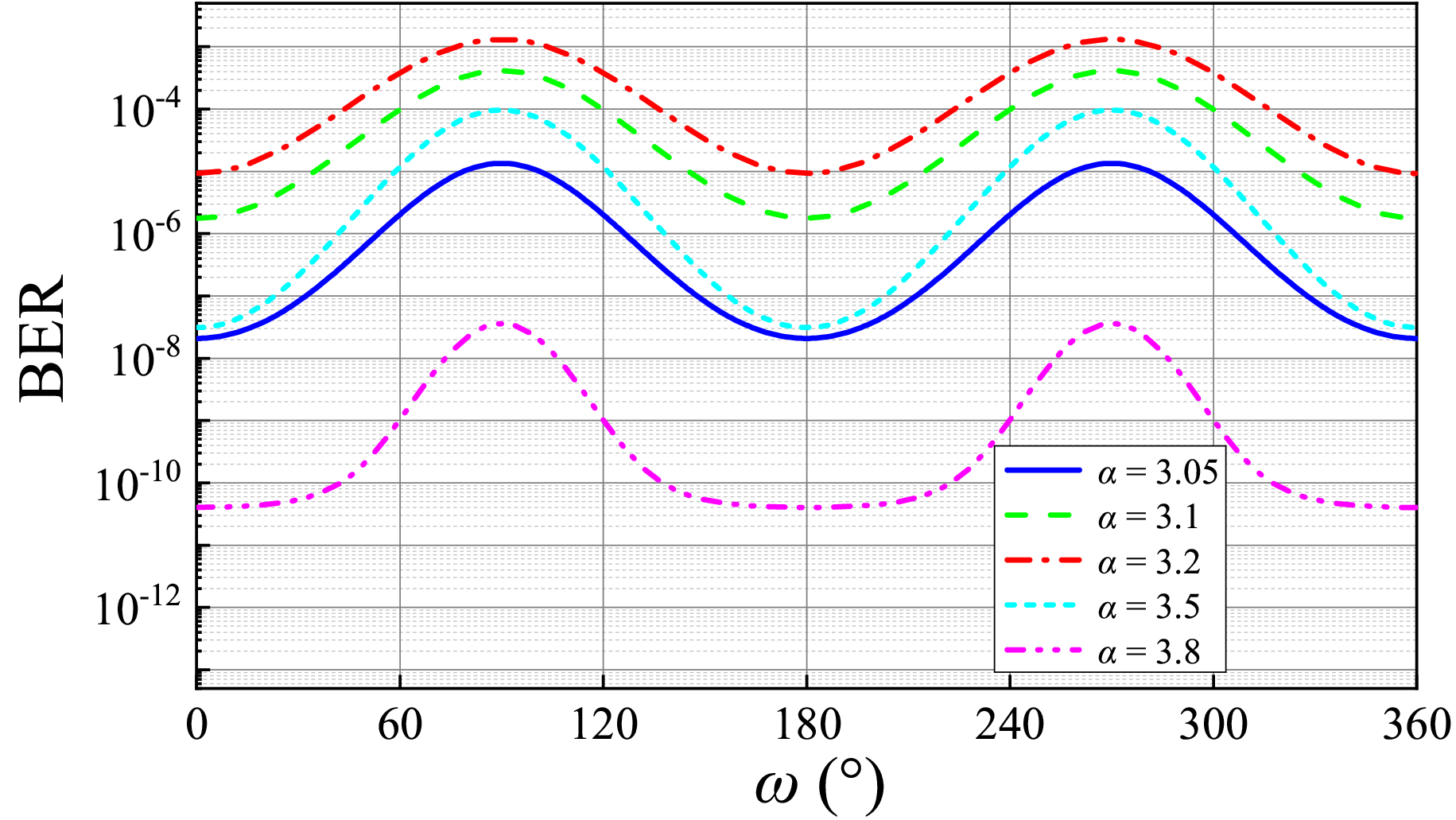}
	\caption{BER of fiber-based QAM/FSO systems under several values of power law $\alpha$ in the horizontal link. (a) as the function of anisotropic tilt angle $\gamma$; (b) as the function of azimuth angle $\omega$.}
	\label{Fig04}
\end{figure}

The average BER of fiber-based QAM/FSO systems for the horizontal link is sketched in Fig.~\ref{Fig04} versus the function of anisotropic tilt angle $\gamma$ or azimuth angle $\omega$ with different power law $\alpha$ values. As displayed in Fig.~\ref{Fig04}, for all the fixed power law $\alpha$ values for any azimuth angle $\omega$ except $ \omega $ = 90 $ ^\circ $ and 270 $ ^\circ $, the average BER of fiber-based QAM/FSO systems first decays and then increases with the growing of anisotropic tilt angle $\gamma$ in the horizontal link, and the variation trends of BER are symmetrical about the straight-line $\gamma$ = 90 $ ^\circ $. And for all the fixed power law $\alpha$ values for any anisotropic tilt angle $\gamma$ except $ \gamma $ = 0 $ ^\circ $ and 180 $ ^\circ $, as the azimuth angle $\omega$ increases from 0 $ ^\circ $ to 180 $ ^\circ $, the average BER of fiber-based QAM/FSO systems increases up to a peak value at $\omega$ = 90 $ ^\circ $ in the horizontal link, and after the peak point, it begins to decay. In addition, the variation trends of BER are symmetrical about the straight-line $\omega$ = 180 $ ^\circ $. We can also observe from Fig.~\ref{Fig04} that, for various values of $\gamma$ and $\omega$, the average BER of fiber-based QAM/FSO systems first rises up to a maximum value around 3.2 to 3.3 and then drops with the enhancement of $\alpha$. This comment can be mainly explained by the fact that, with the growing of $\alpha$, the scintillation index of a plane wave first increases up to a maximum value around 3.2 to 3.3 and then decays in the ANK horizontal link \cite{Andrews2014}. An increase in the scintillation index means that the effect of turbulence on the propagated beam will be enhanced, then the average BER of fiber-based QAM/FSO systems will increase accordingly.

\begin{figure}[ht!]
	\centering
	\includegraphics[width=6.6cm]{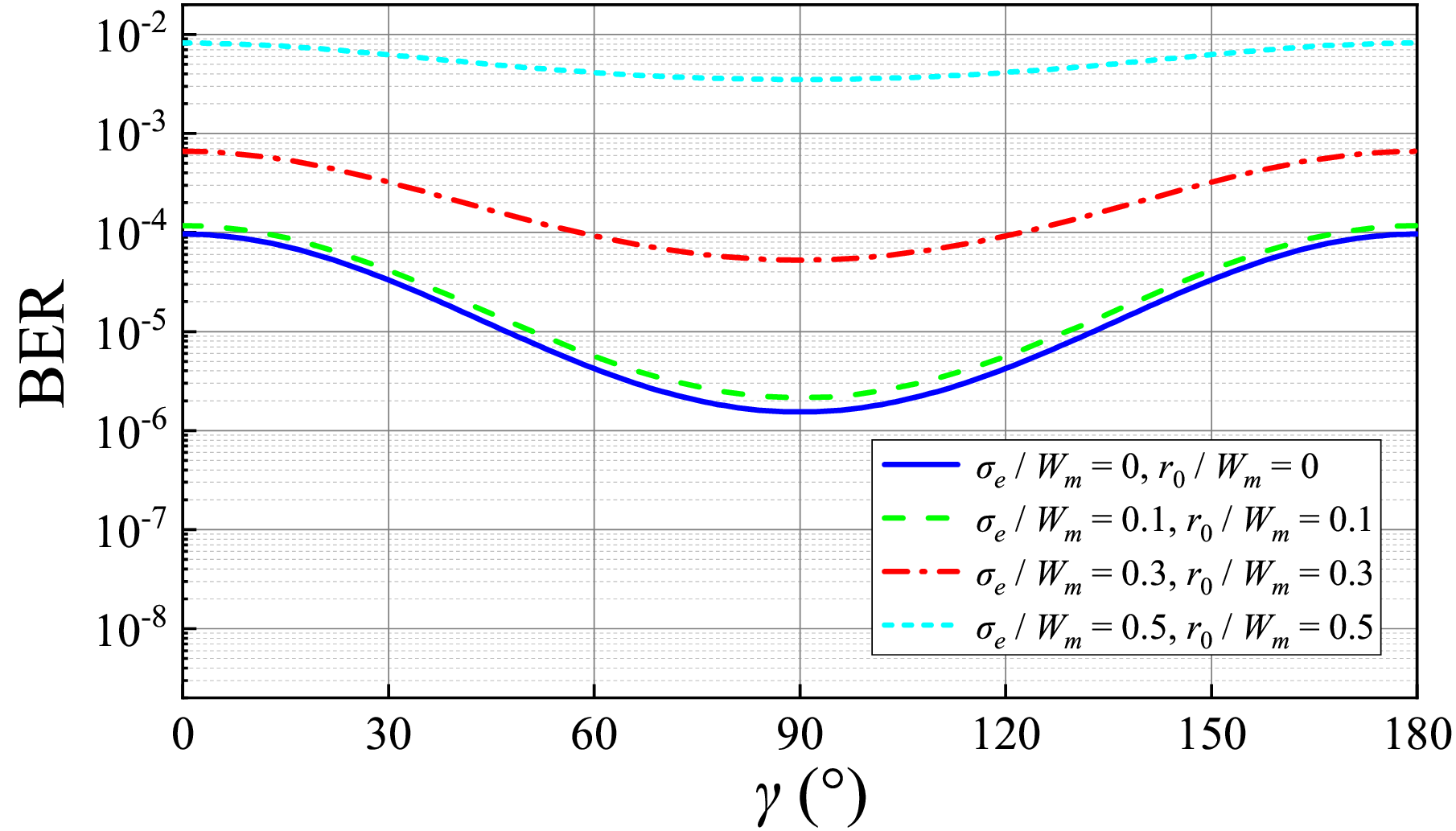}
	\includegraphics[width=6.6cm]{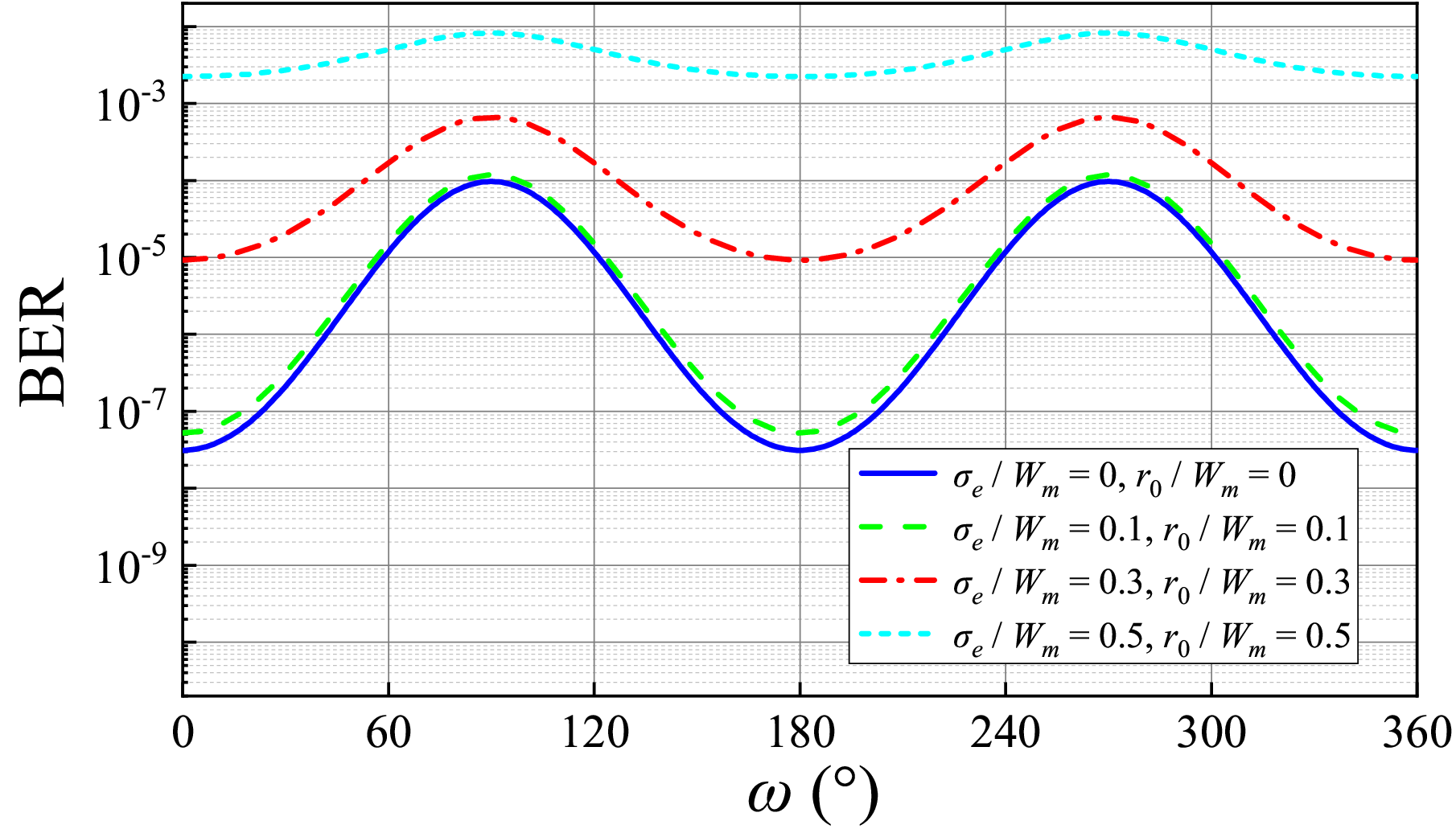}
	\caption{BER of fiber-based QAM/FSO systems under several values of normalized random angular jitter ${\sigma _e}/{W_m}$ and normalized bias error ${r_0}/{W_m}$ in the horizontal link. (a) as the function of anisotropic tilt angle $\gamma$; (b) as the function of azimuth angle $\omega$.}
	\label{Fig05}
\end{figure}

To fully reveal the impacts of the normalized random angular jitter ${\sigma _e}/{W_m}$, the anisotropic tilt angle $\gamma$, the normalized bias error ${r_0}/{W_m}$, and the azimuth angle $\omega$ on the average BER of fiber-based QAM/FSO systems in the horizontal link, the variation of BER against $\gamma$ or $\omega$ with different values of ${\sigma _e}/{W_m}$ and ${r_0}/{W_m}$ is plotted in Fig.~\ref{Fig05}. As displayed in Fig.~\ref{Fig05}, for all the fixed values of normalized random angular jitter ${\sigma _e}/{W_m}$ and normalized bias error ${r_0}/{W_m}$ for any azimuth angle $\omega$ except $ \omega $ = 90 $ ^\circ $ and 270 $ ^\circ $, the average BER of fiber-based QAM/FSO systems first decays and then increases with the growing of anisotropic tilt angle $\gamma$ in the horizontal link, and the variation trends of BER are symmetrical about the straight-line $\gamma$ = 90 $ ^\circ $. And for all the fixed values of normalized random angular jitter ${\sigma _e}/{W_m}$ and normalized bias error ${r_0}/{W_m}$ for any anisotropic tilt angle $\gamma$ except $ \gamma $ = 0 $ ^\circ $ and 180 $ ^\circ $, as the azimuth angle $\omega$ increases from 0 $ ^\circ $ to 180 $ ^\circ $, the average BER of fiber-based QAM/FSO systems increases up to a peak value at $\omega$ = 90 $ ^\circ $ in the horizontal link, and after the peak point, it begins to decay. In addition, the variation trends of BER are symmetrical about the straight-line $\omega$ = 180 $ ^\circ $. It can also be deduced from Fig.~\ref{Fig05} that, for various values of $\gamma$ and $\omega$, the average BER of fiber-based QAM/FSO systems increases with growing of ${\sigma _e}/{W_m}$ and ${r_0}/{W_m}$. As shown in Fig.~\ref{Fig02}, due to the unpredictability between the instantaneous direction of received laser and the nominal axis of fiber, bias error $ r_0 $ and random angular jitter $ \sigma _e $ are generated, which will result in a lateral displacement $ \Delta r $ and a static radial displacement between the nominal axis of fiber and the optical axis of receiver lens at the focal plane, respectively. Because of the lateral displacement $ \Delta r $ and the static radial displacement, the matching degree of the backpropagated fiber-mode profile $ {U_m}\left( {\bf{r}} \right) $ and the incident optical field $ {U_i}\left( {\bf{r}} \right) $ will be reduced, and the fiber-coupling efficiency will decrease accordingly, finally bring down the performance of FSO communication systems. Thus, in the process of modeling and analysis of the practical fiber-based QAM/FSO systems, we must evaluate the influences of bias error $ r_0 $ and random angular jitter $ \sigma _e $ comprehensively.

\begin{figure}[ht!]
	\centering
	\includegraphics[width=6.6cm]{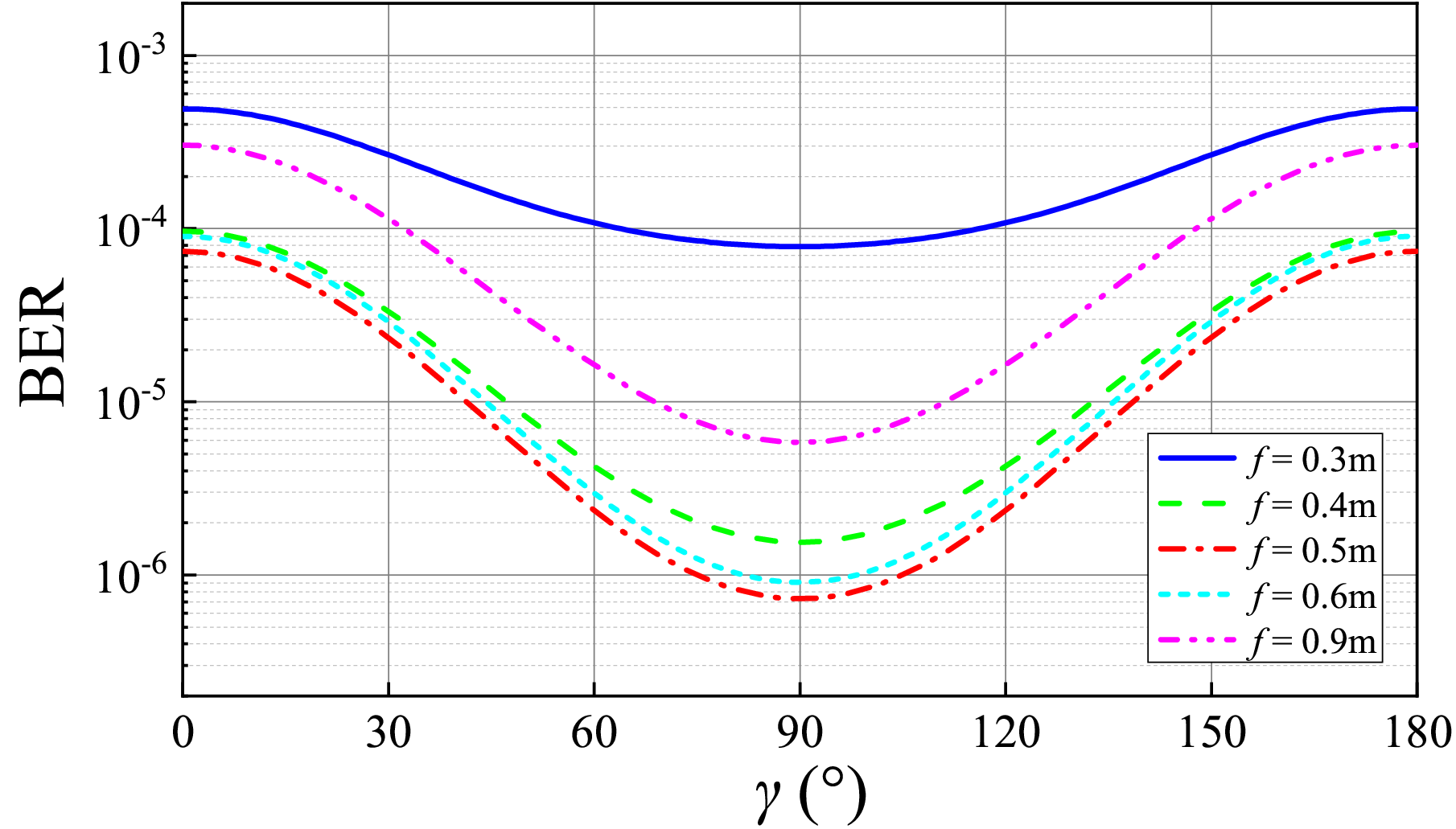}
	\includegraphics[width=6.6cm]{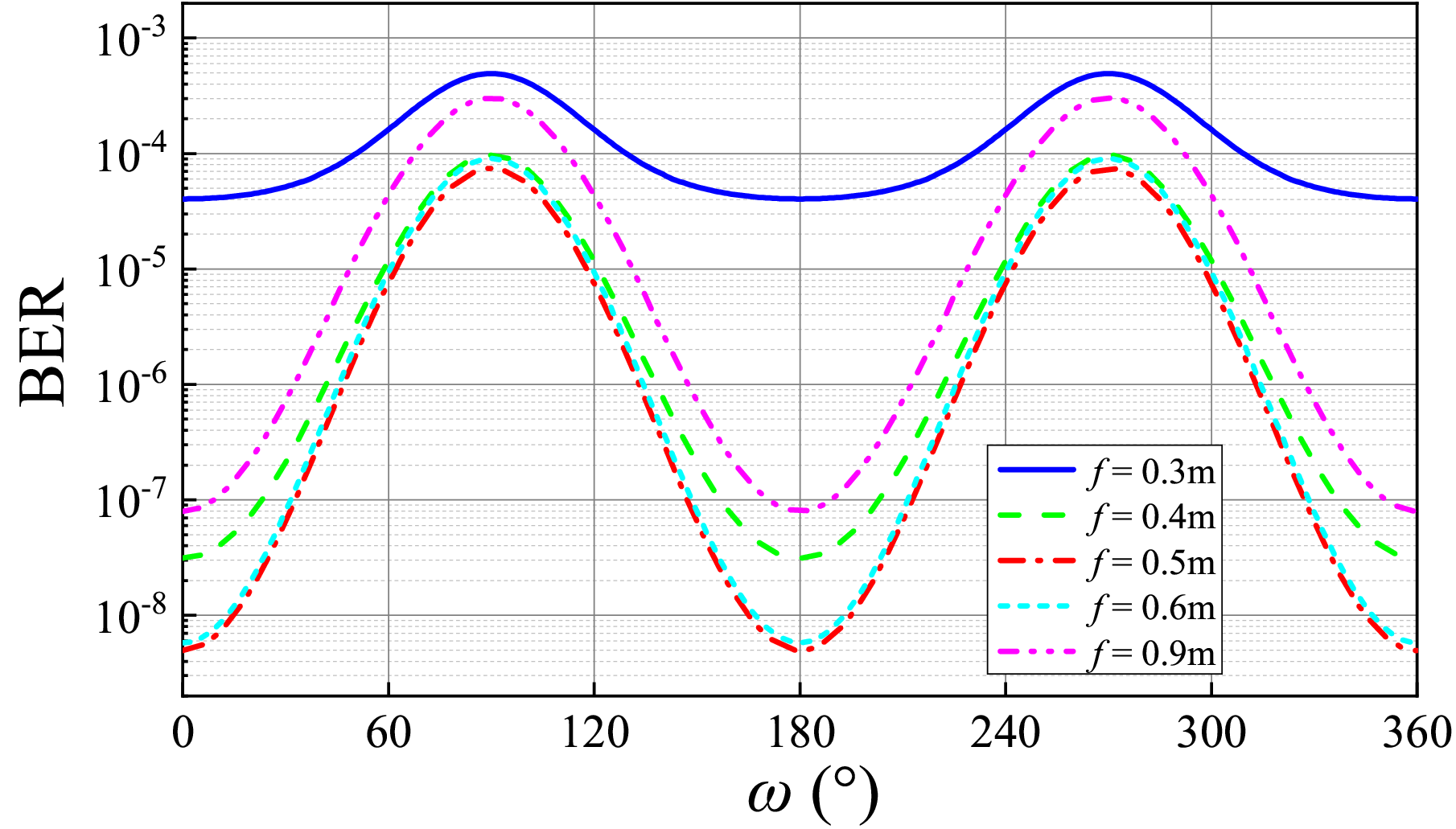}
	\caption{BER of fiber-based QAM/FSO systems under several values of focal length $ f $ in the horizontal link. (a) as the function of anisotropic tilt angle $\gamma$; (b) as the function of azimuth angle $\omega$.}
	\label{Fig06}
\end{figure}

Fig.~\ref{Fig06} presents the deviation of average BER of fiber-based QAM/FSO systems in the horizontal link as the function of anisotropic tilt angle $\gamma$ or azimuth angle $\omega$ under several values of focal length $ f $. As displayed in Fig.~\ref{Fig06}, for all the fixed focal length $ f $ values for any azimuth angle $\omega$ except $ \omega $ = 90 $ ^\circ $ and 270 $ ^\circ $, the average BER of fiber-based QAM/FSO systems first decays and then increases with the growing of anisotropic tilt angle $\gamma$ in the horizontal link, and the variation trends of BER are symmetrical about the straight-line $\gamma$ = 90 $ ^\circ $. And for all the fixed focal length $ f $ values for any anisotropic tilt angle $\gamma$ except $ \gamma $ = 0 $ ^\circ $ and 180 $ ^\circ $, as the azimuth angle $\omega$ increases from 0 $ ^\circ $ to 180 $ ^\circ $, the average BER of fiber-based QAM/FSO systems increases up to a peak value at $\omega$ = 90 $ ^\circ $ in the horizontal link, and after the peak point, it begins to decay. In addition, the variation trends of BER are symmetrical about the straight-line $\omega$ = 180 $ ^\circ $. It can also be deduced from Fig.~\ref{Fig06} that, for various values of $\gamma$ and $\omega$, the average BER of fiber-based QAM/FSO systems first decays and then rises up with the enhancement of $ f $, and we can obtain an optimal $ f $ that maximizes the BER performance of FSO communication systems. Physically, the fiber-coupling efficiency represents the matching degree of the backpropagated fiber-mode profile $ {U_m}\left( {\bf{r}} \right) $ and the incident optical field $ {U_i}\left( {\bf{r}} \right) $. When the receiver diameter $D$ is fixed, we can easily get that the radius of $ {U_i}\left( {\bf{r}} \right) $ is also fixed. Then for several fixed values of wavelength $ \lambda $ and fiber-mode field radius $ W_m $, an optimal focal length $ f $ can be found to maximize the matching degree. Accordingly, an optimal fiber-coupling efficiency will lead to the better BER performance of FSO communication systems.

\begin{figure}[ht!]
	\centering
	\includegraphics[width=6.6cm]{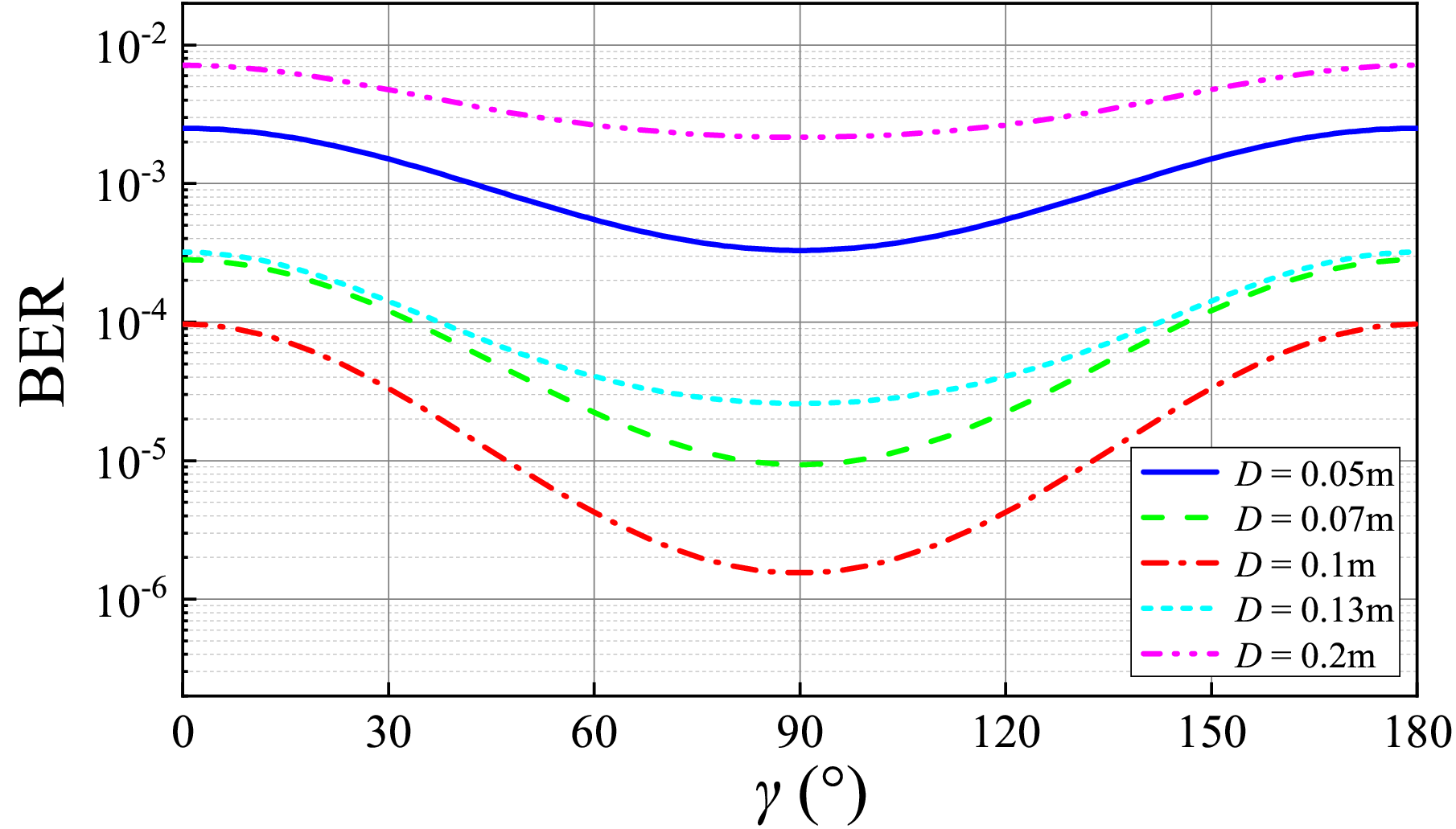}
	\includegraphics[width=6.6cm]{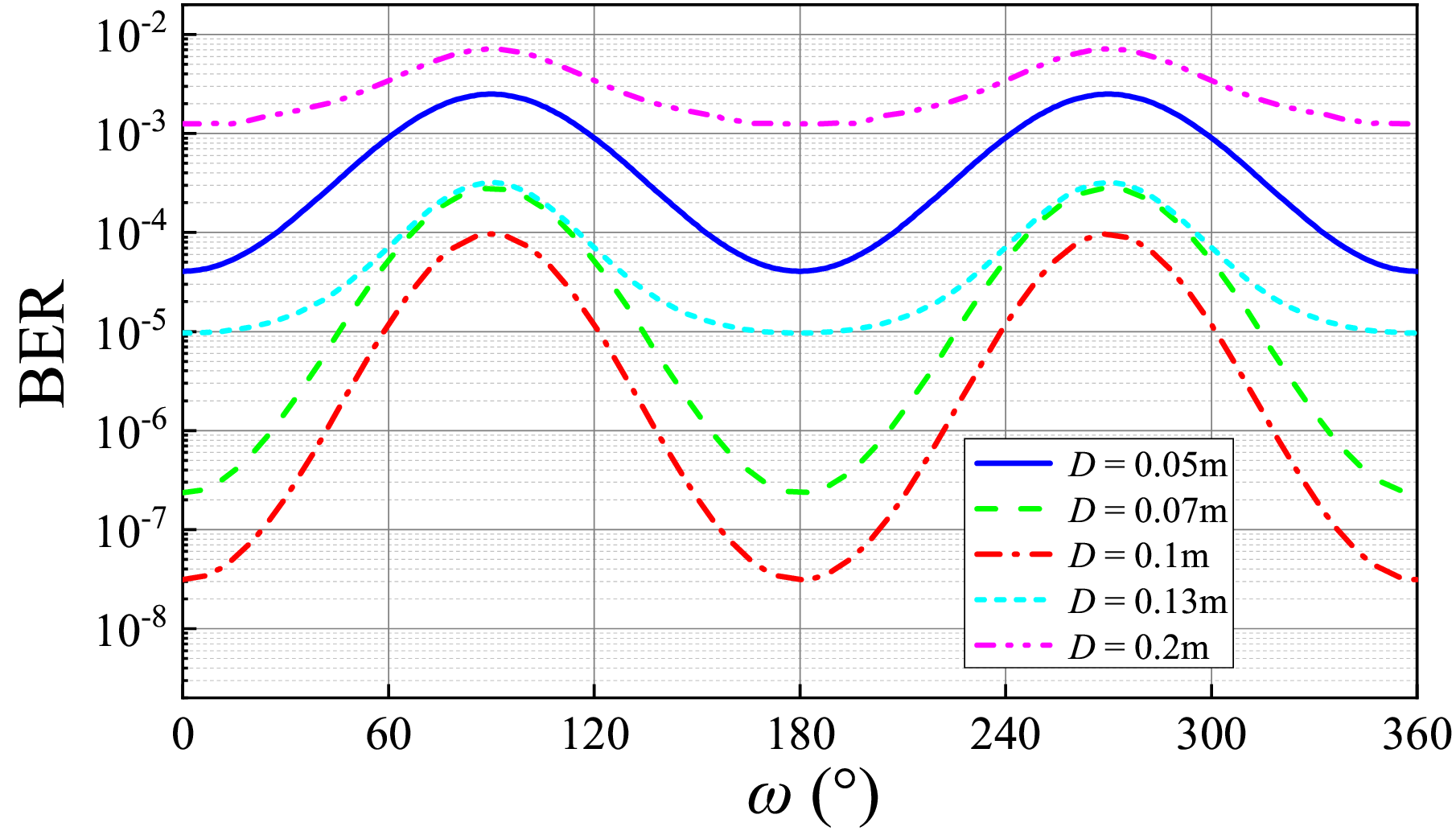}
	\caption{BER of fiber-based QAM/FSO systems under several values of receiver diameter $D$ in the horizontal link. (a) as the function of anisotropic tilt angle $\gamma$; (b) as the function of azimuth angle $\omega$.}
	\label{Fig07}
\end{figure}

In Fig.~\ref{Fig07}, the average BER of fiber-based QAM/FSO systems in the horizontal link is drawn versus the function of anisotropic tilt angle $\gamma$ or azimuth angle $\omega$ at different receiver diameter $D$ values. As demonstrated in Fig.~\ref{Fig07}, for all the fixed receiver diameter $D$ values for any azimuth angle $\omega$ except $ \omega $ = 90 $ ^\circ $ and 270 $ ^\circ $, the average BER of fiber-based QAM/FSO systems first decays and then increases with the growing of anisotropic tilt angle $\gamma$ in the horizontal link, and the variation trends of BER are symmetrical about the straight-line $\gamma$ = 90 $ ^\circ $. And for all the fixed receiver diameter $D$ values for any anisotropic tilt angle $\gamma$ except $ \gamma $ = 0 $ ^\circ $ and 180 $ ^\circ $, as the azimuth angle $\omega$ increases from 0 $ ^\circ $ to 180 $ ^\circ $, the average BER of fiber-based QAM/FSO systems increases up to a peak value at $\omega$ = 90 $ ^\circ $ in the horizontal link, and after the peak point, it begins to decay. In addition, the variation trends of BER are symmetrical about the straight-line $\omega$ = 180 $ ^\circ $. We can also obtain from Fig.~\ref{Fig07} that, for various values of $\gamma$ and $\omega$, as the value of $D$ increases, the average BER of fiber-based QAM/FSO systems first decays and then rises up, and an optimal $D$ can be found to maximize the BER performance of FSO communication systems. This comment can be clarified from a physical idea that the fiber-coupling efficiency relies on the matching degree of $ {U_i}\left( {\bf{r}} \right) $ and $ {U_m}\left( {\bf{r}} \right) $. When the focal length $ f $, the wavelength $ \lambda $, and the fiber-mode field radius $ W_m $ are all fixed, the radius of $ {U_m}\left( {\bf{r}} \right) $ is also fixed. Hence, there will be an appropriate $D$ to optimize the matching degree. Meanwhile, as the receiver diameter $D$ increases, the SNR per bit $ \gamma _b $ increases, bringing about a decrease in the average BER of fiber-based QAM/FSO systems. Finally, the fiber-coupling efficiency has a larger impact on the average BER of fiber-based QAM/FSO systems, thus we can obtain an optimal $D$ to make the BER performance of FSO communication systems better.

\begin{figure}[ht!]
	\centering
	\includegraphics[width=6.6cm]{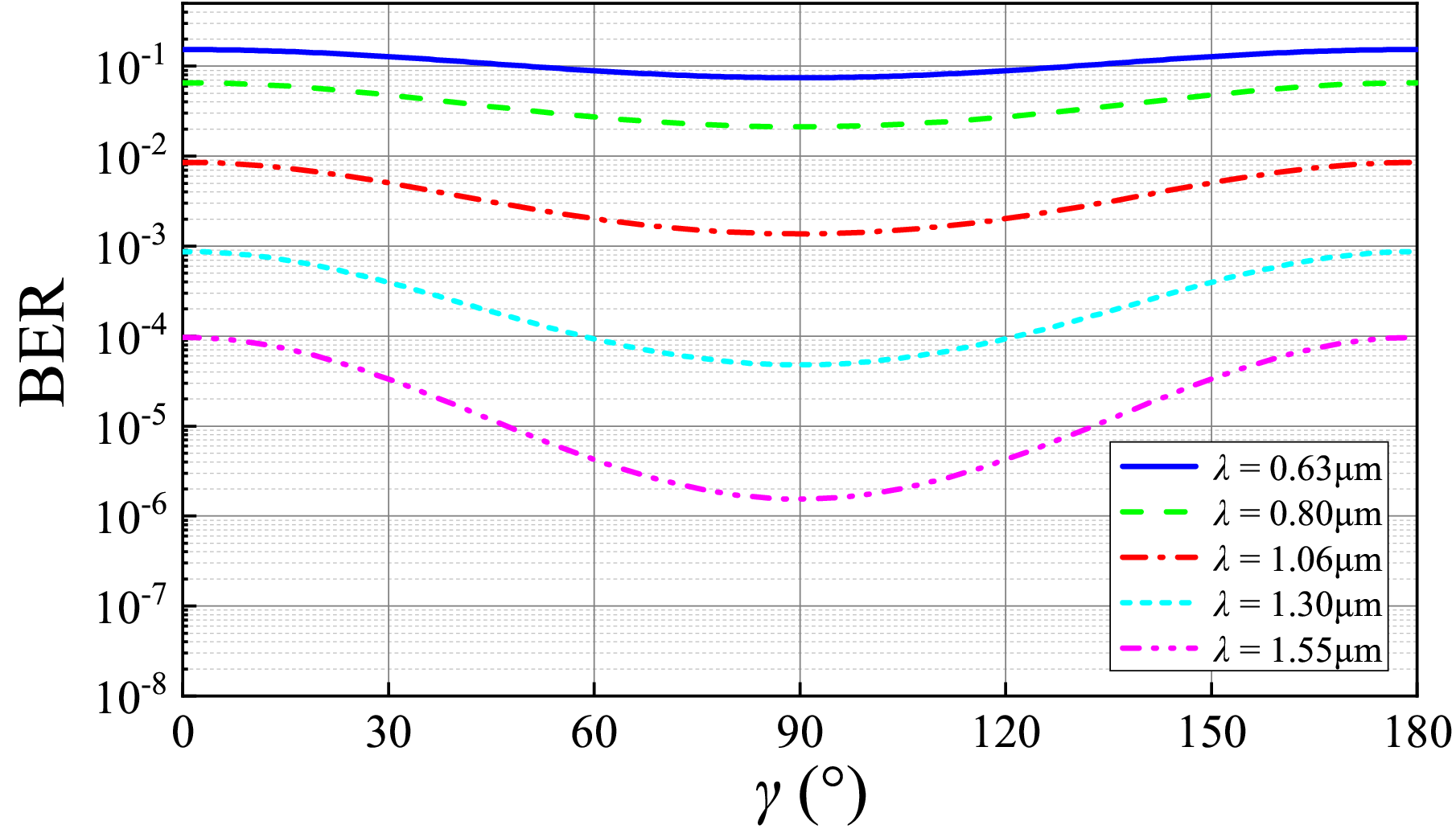}
	\includegraphics[width=6.6cm]{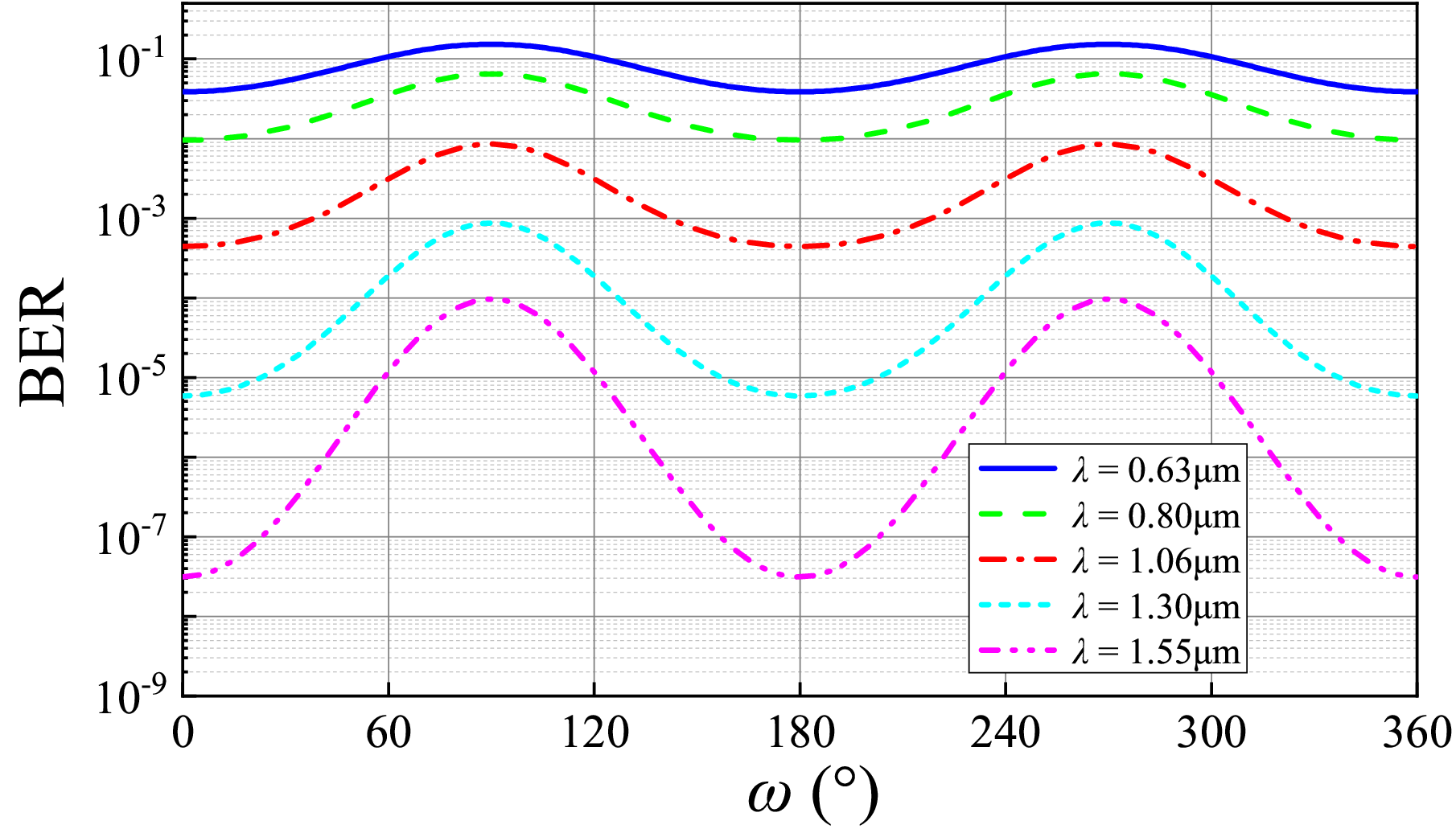}
	\caption{BER of fiber-based QAM/FSO systems under several values of wavelength $ \lambda $ in the horizontal link. (a) as the function of anisotropic tilt angle $\gamma$; (b) as the function of azimuth angle $\omega$.}
	\label{Fig08}
\end{figure}

Figure~\ref{Fig08} displays the variation of average BER of fiber-based QAM/FSO systems in the horizontal link against the function of anisotropic tilt angle $\gamma$ or azimuth angle $\omega$ with some fixed values of wavelength $ \lambda $. As displayed in Fig.~\ref{Fig08}, for all the fixed wavelength $ \lambda $ values for any azimuth angle $\omega$ except $ \omega $ = 90 $ ^\circ $ and 270 $ ^\circ $, the average BER of fiber-based QAM/FSO systems first decays and then increases with the growing of anisotropic tilt angle $\gamma$ in the horizontal link, and the variation trends of BER are symmetric about the straight-line $\gamma$ = 90 $ ^\circ $. And for all the fixed wavelength $ \lambda $ values for any anisotropic tilt angle $\gamma$ except $ \gamma $ = 0 $ ^\circ $ and 180 $ ^\circ $, as the azimuth angle $\omega$ increases from 0 $ ^\circ $ to 180 $ ^\circ $, the average BER of fiber-based QAM/FSO systems increases up to a peak value at $\omega$ = 90 $ ^\circ $ in the horizontal link, and after the peak point, it begins to decay. In addition, the variation trends of BER are symmetric about the straight-line $\omega$ = 180 $ ^\circ $. It can also be observed from Fig.~\ref{Fig08} that, for various values of $\gamma$ and $\omega$, the average BER of fiber-based QAM/FSO systems decays with the enhancement of $\lambda $. This comment is consistent with our former research, which indicated that, as the $ \lambda $ became larger, the impact of turbulence on the propagated laser beam would be weakened \cite{Zhai2022}.

\begin{figure}[ht!]
	\centering
	\includegraphics[width=6.6cm]{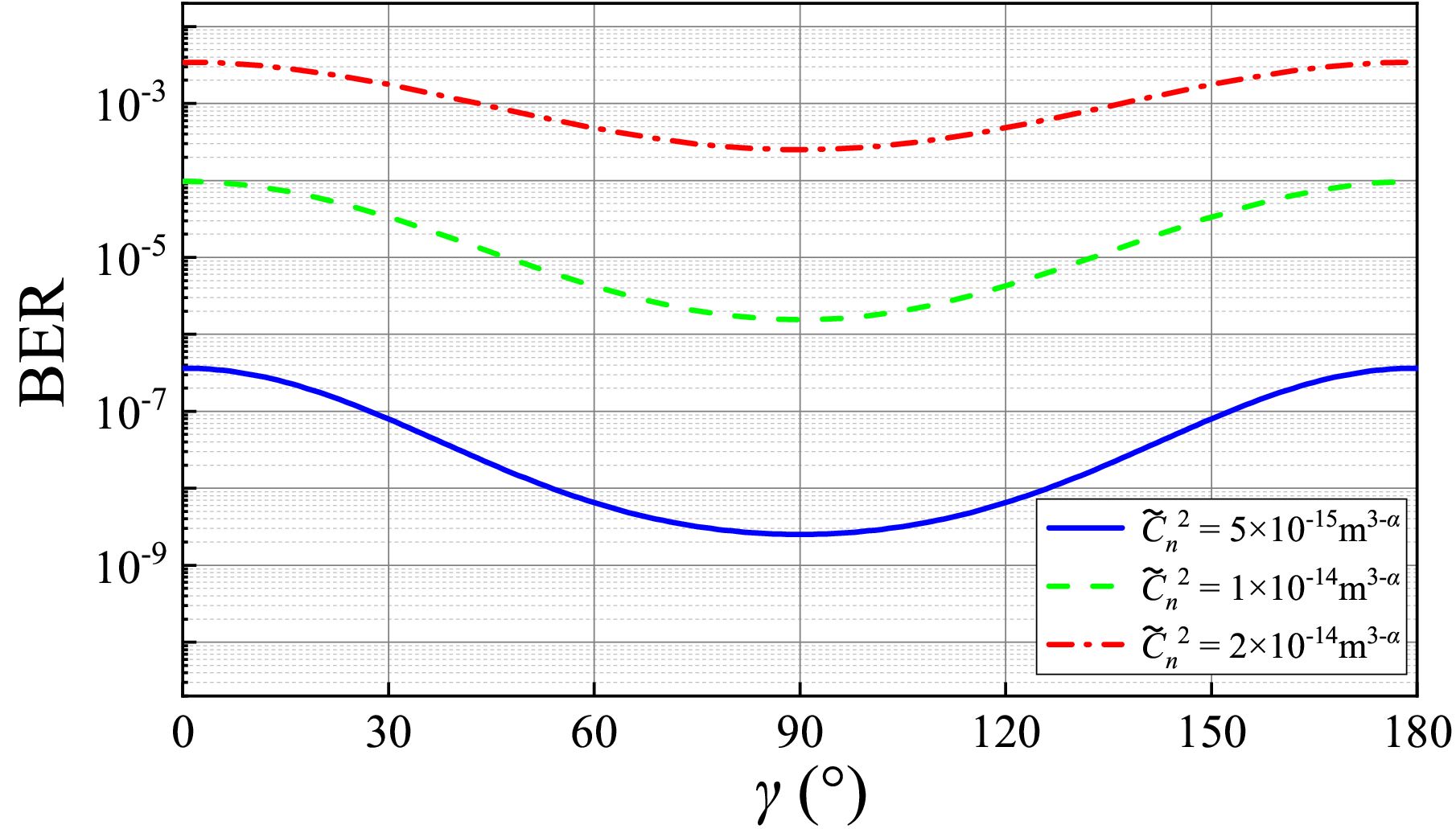}
	\includegraphics[width=6.6cm]{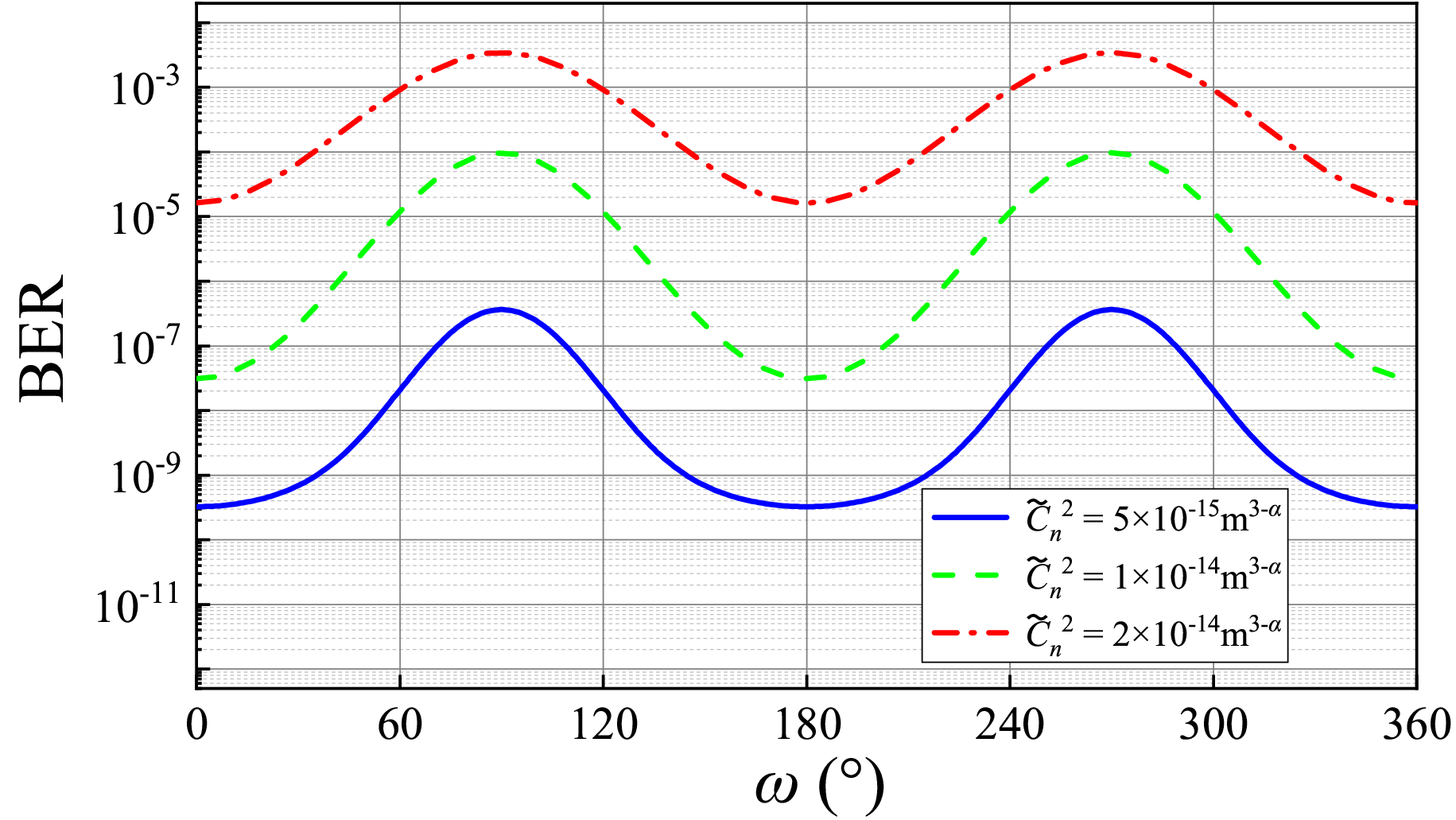}
	\caption{BER of fiber-based QAM/FSO systems under several values of turbulence structure constant $\tilde C_n^2$ in the horizontal link. (a) as the function of anisotropic tilt angle $\gamma$; (b) as the function of azimuth angle $\omega$.}
	\label{Fig09}
\end{figure}

To further examine the influences of the anisotropic tilt angle $\gamma$, the turbulence structure constant $\tilde C_n^2$, and the azimuth angle $\omega$ on the average BER of fiber-based QAM/FSO systems in the horizontal link, the BER versus $\gamma$ or $\omega$ under several values of $\tilde C_n^2$ is demonstrated in Fig.~\ref{Fig09}. As displayed in Fig.~\ref{Fig09}, for all the fixed values of turbulence structure constant $\tilde C_n^2$ for any azimuth angle $\omega$ except $ \omega $ = 90 $ ^\circ $ and 270 $ ^\circ $, the average BER of fiber-based QAM/FSO systems first decays and then increases with the growing of anisotropic tilt angle $\gamma$ in the horizontal link, and the variation trends of BER are symmetrical about the straight-line $\gamma$ = 90 $ ^\circ $. And for all the fixed values of turbulence structure constant $\tilde C_n^2$ for any anisotropic tilt angle $\gamma$ except $ \gamma $ = 0 $ ^\circ $ and 180 $ ^\circ $, as the azimuth angle $\omega$ increases from 0 $ ^\circ $ to 180 $ ^\circ $, the average BER of fiber-based QAM/FSO systems increases up to a peak value at $\omega$ = 90 $ ^\circ $ in the horizontal link, and after the peak point, it begins to decay. In addition, the variation trends of BER are symmetrical about the straight-line $\omega$ = 180 $ ^\circ $. We can also observe from Fig.~\ref{Fig09} that, for various values of $\gamma$ and $\omega$, an increment in $\tilde C_n^2$ will bring about an increase in the average BER of fiber-based QAM/FSO systems. This comment can be interpreted by the fact that larger turbulence structure constant $\tilde C_n^2$ would yield smaller fiber-coupling efficiency and larger scintillation index \cite{Zhai2020,Zhai2022}, and then the average BER of fiber-based QAM/FSO systems would increase accordingly.

\begin{figure}[ht!]
	\centering
	\includegraphics[width=6.6cm]{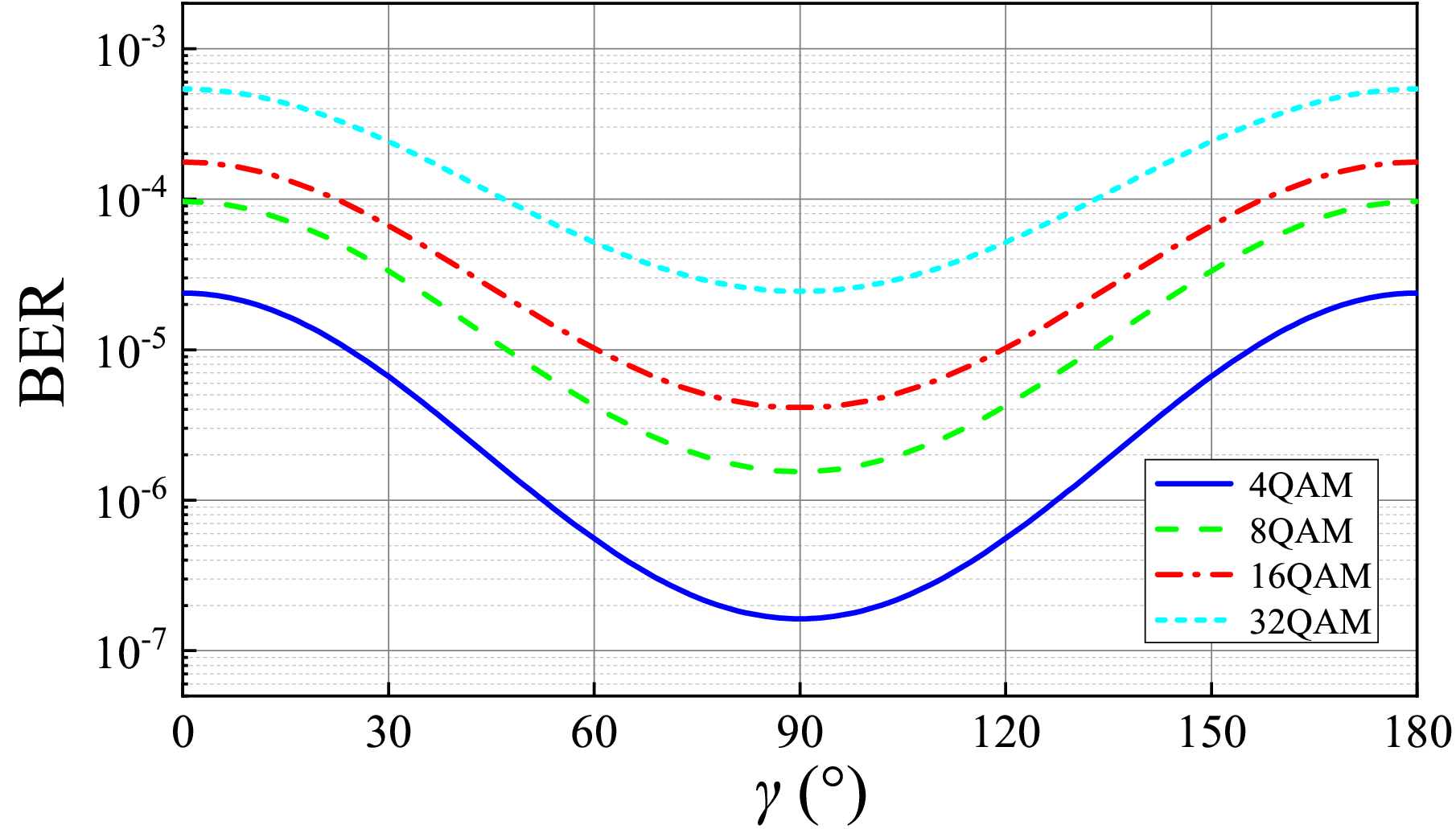}
	\includegraphics[width=6.6cm]{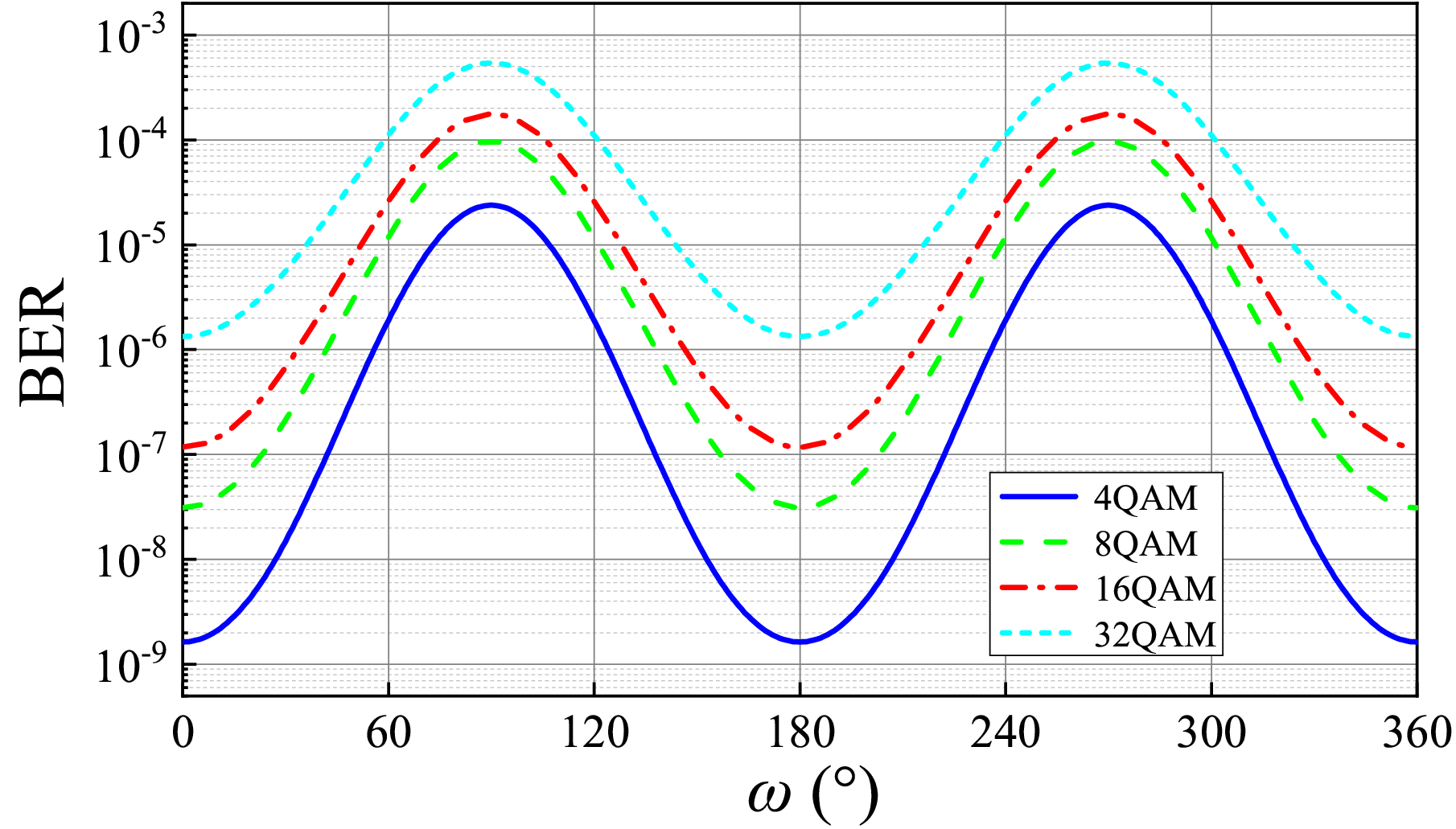}
	\caption{BER of fiber-based QAM/FSO systems under different QAM schemes in the horizontal link. (a) as the function of anisotropic tilt angle $\gamma$; (b) as the function of azimuth angle $\omega$.}
	\label{Fig10}
\end{figure}
\begin{figure}[ht!]
	\centering
	\includegraphics[width=6.6cm]{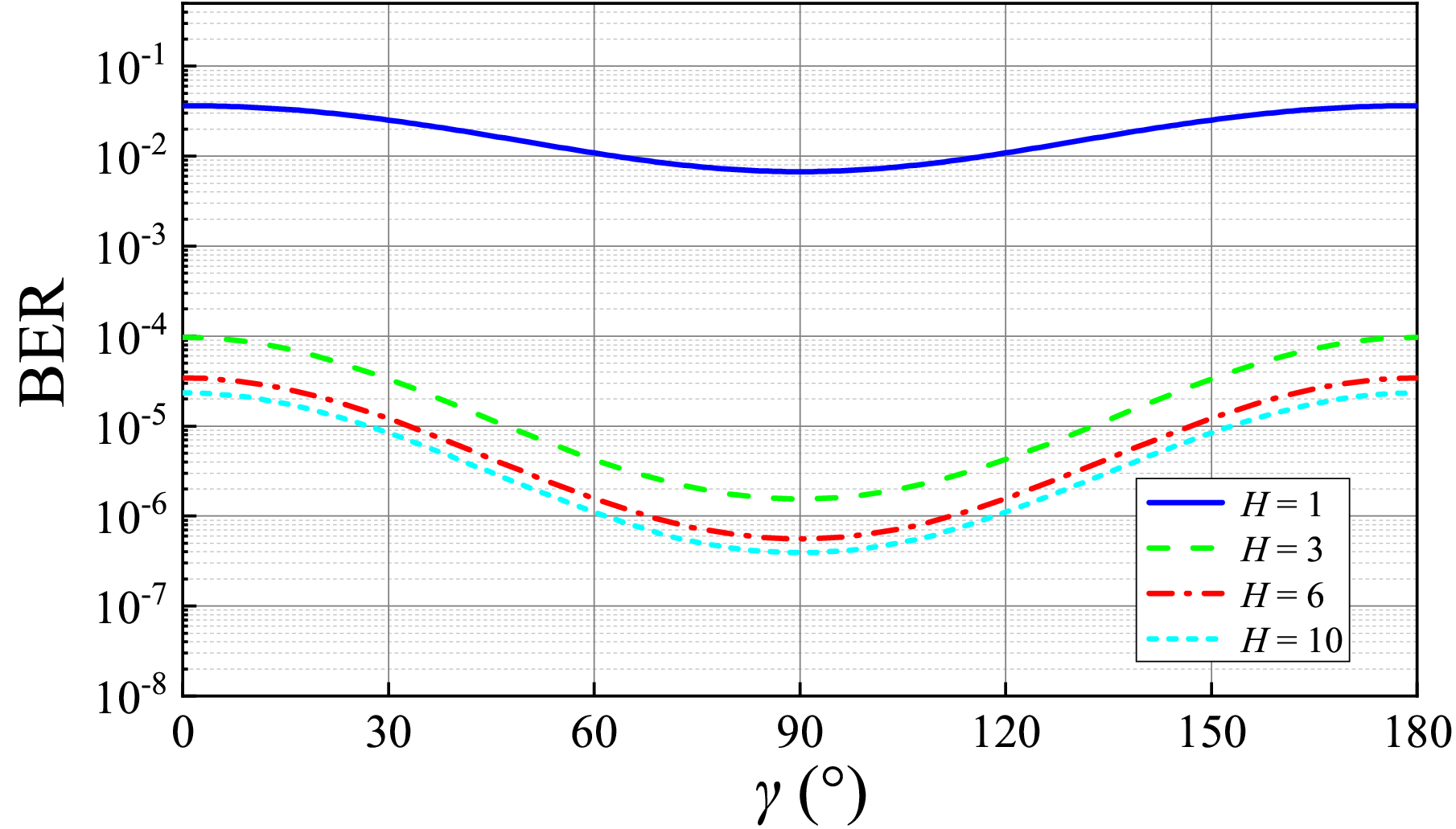}
	\includegraphics[width=6.6cm]{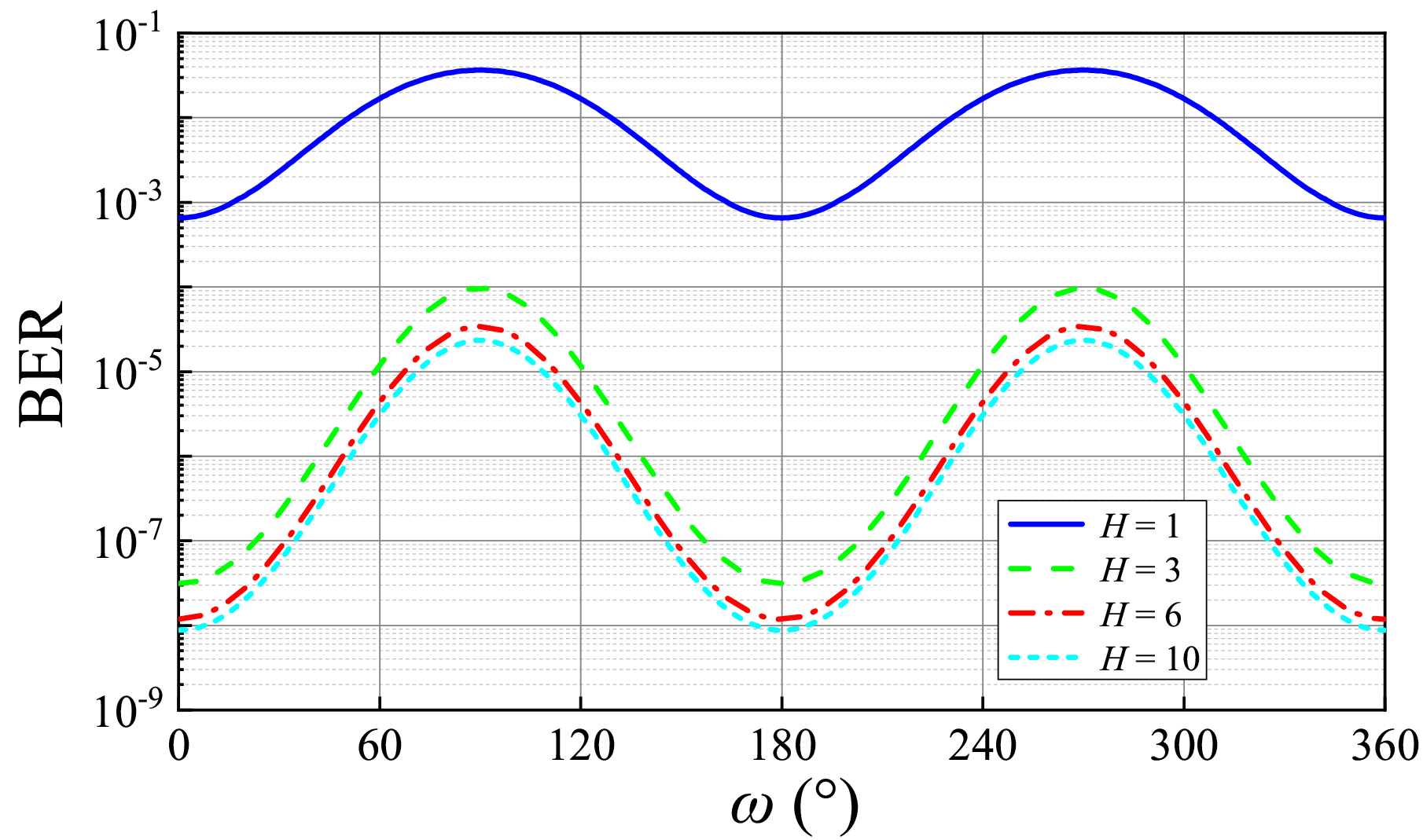}
	\caption{BER of fiber-based QAM/FSO systems under several numbers of compensation terms $H$ in the horizontal link. (a) as the function of anisotropic tilt angle $\gamma$; (b) as the function of azimuth angle $\omega$.}
	\label{Fig11}
\end{figure}

To clarify the impacts of the QAM scheme, the anisotropic tilt angle $\gamma$, the number of compensation terms $H$, and the azimuth angle $\omega$ on the average BER of fiber-based QAM/FSO systems in the horizontal link, the variation of BER against $\gamma$ or $\omega$ under different QAM schemes ($ J \times Q = 2 \times 2 $, $ 4 \times 2 $, $ 4 \times 4 $, and $ 8 \times 4 $) is exhibited in Fig.~\ref{Fig10}, and the variation of BER against $\gamma$ or $\omega$ under several values of $H$ is exhibited in Fig.~\ref{Fig11}. As demonstrated in Figs.~\ref{Fig10} and~\ref{Fig11}, for all the QAM schemes and the numbers of compensation terms $H$ for any azimuth angle $\omega$ except $ \omega $ = 90 $ ^\circ $ and 270 $ ^\circ $, the average BER of fiber-based QAM/FSO systems first decays and then increases with the growing of anisotropic tilt angle $\gamma$ in the horizontal link, and the variation trends of BER are symmetrical about the straight-line $\gamma$ = 90 $ ^\circ $. And for all the QAM schemes and the numbers of compensation terms $H$ for any anisotropic tilt angle $\gamma$ except $ \gamma $ = 0 $ ^\circ $ and 180 $ ^\circ $, as the azimuth angle $\omega$ increases from 0 $ ^\circ $ to 180 $ ^\circ $, the average BER of fiber-based QAM/FSO systems increases up to a peak value at $\omega$ = 90 $ ^\circ $ in the horizontal link, and after the peak point, it begins to decay. In addition, the variation trends of BER are symmetrical about the straight-line $\omega$ = 180 $ ^\circ $. It can also be observed from Figs.~\ref{Fig10} and~\ref{Fig11} that, for various values of $\gamma$ and $\omega$, FSO communication systems have worse BER performance with a higher QAM scheme, but a larger quantity of information can be sent, and the average BER of fiber-based QAM/FSO systems decays with the enhancement of $H$. This comment can be interpreted by the fact that, as the constellation size of $J \times Q$ rectangular QAM increases, the SNR per bit $ \gamma _b $ decays, bringing about an increase in the average BER of fiber-based QAM/FSO systems. Moreover, as the number of compensation terms $H$ increases, the residual phase variance after phase compensation $ \sigma _\phi ^2 $ will decay, resulting in the improvement of fiber-coupling efficiency, and then the average BER of fiber-based QAM/FSO systems will decay accordingly.

\section{Conclusion}
With the aid of new ANK turbulence spectrum models in the horizontal link with anisotropic tilt angle, we develop the average BER expression of fiber-based $J \times Q$ rectangular QAM/FSO systems for a plane wave transmission along the weak ANK horizontal link in the presence of bias error, anisotropic tilt angle, and random angular jitter. The numerical calculation results indicate that for any azimuth angle $\omega$ except $ \omega $ = 90 $ ^\circ $ and 270 $ ^\circ $, the average BER of fiber-based QAM/FSO systems first decays and then increases with the growing of anisotropic tilt angle $\gamma$, and the variation trends of BER are symmetrical about the straight-line $\gamma$ = 90 $ ^\circ $. Similarly, for any anisotropic tilt angle $\gamma$ except $ \gamma $ = 0 $ ^\circ $ and 180 $ ^\circ $, as the azimuth angle $\omega$ increases from 0 $ ^\circ $ to 180 $ ^\circ $, the average BER of fiber-based QAM/FSO systems increases up to a peak value at $\omega$ = 90 $ ^\circ $, and after the peak point, it begins to decay. And the variation trends of BER are symmetrical about the straight-line $\omega$ = 180 $ ^\circ $. In the region that is closer to the plane where the long axes of turbulence cell ellipsoid model are located, the average BER of fiber-based QAM/FSO systems increases with the enhancement of anisotropic factor $\mu$. But for the region that is closer to the short axis of turbulence cell ellipsoid model, the contrary trend will occur. For the horizontal link, an increase for the wavelength, the number of compensation terms, and a decrease for the random angular jitter, the bias error, the turbulence structure constant, the QAM scheme will result in the better BER performance of FSO communication systems. And with the enhancement of power law, the average BER of fiber-based QAM/FSO systems first rises up to a maximum value around 3.2 to 3.3 and then drops. In addition, the average BER of fiber-based QAM/FSO systems first decays and then rises up with the enhancement of focal length $ f $ and receiver diameter $ D $, and the optimum $ f $ and $ D $ can be obtained to improve the BER performance of FSO communication systems. Our research will be advantageous to the design improvement of fiber-based QAM/FSO communication systems in the ANK horizontal link with anisotropic tilt angle.

\section{CRediT authorship contribution statement}
\textbf{Chao Zhai:} Writing - original draft, Writing - review $\&$ editing, Conceptualization, Methodology, Software, Validation, Supervision. \textbf{Zhenyuan Xue:} Writing - review $\&$ editing, Investigation, Validation.

\section{Declaration of competing interest}
The authors declare that they have no known competing financial interests or personal relationships that could have appeared to influence the work reported in this paper.

\section{Acknowledgements}
The authors gratefully acknowledge the financial support by the National Natural Science Foundation of China (61901097).

\section{Data availability}
Data will be made available on request.

\bibliography{Lib}
	
\end{document}